\documentclass[doublespacing,preprint,aip,jcp]{revtex4-1}
\usepackage{graphicx}
\usepackage{bm}
\usepackage{amsmath}
\DeclareMathOperator{\sgn}{sgn}
\usepackage{amssymb}
\usepackage{subfigure}
\usepackage{gensymb}
\usepackage{epstopdf}
\usepackage{hyphenat}
\usepackage[usenames, dvipsnames]{color}

\usepackage[normalem]{ulem}
\usepackage{comment}
\usepackage{csquotes}
\begin{document}

\title{Kosmotropic effect leads to LCST decrease in thermoresponsive polymer solutions }
\author{Swaminath Bharadwaj}
 \affiliation{Department of Chemical Engineering, Indian Institute of Technology Madras, Chennai-600036, India}
\author{P. B. Sunil Kumar}
 \email{sunil@iitpkd.ac.in}
\affiliation{ 
Department of Physics, Indian Institute of Technology Palakkad, Ahalia Integrated Campus, Kozhippara, Palakkad-678557, India}
\author{Shigeyuki Komura}
 \affiliation{Department of Chemistry, Graduate School of Science and Engineering, Tokyo Metropolitan University, Tokyo 192-0397, Japan}
\author{Abhijit P. Deshpande}
 \affiliation{Department of Chemical Engineering, Indian Institute of Technology Madras, Chennai-600036, India}
\date{\today}
\begin{abstract}
We study the phenomena of decrease in lower critical solution temperature (LCST) with addition of kosmotropic (order-making) cosolvents in thermoresponsive polymer solutions. A combination of explicit solvent coarse-grained simulations and  mean-field theory has been employed. The polymer-solvent LCST behavior  in the theoretical models have been incorporated through the Kolomeisky-Widom solvophobic potential. Our results illustrate how  the decrease in the LCST  can be achieved by  the reduction in the bulk solvent energy with addition of cosolvent. It is shown that this effect of cosolvent is weaker with increase in  polymer hydrophilicity which   can  explain the absence of LCST decrease in PDEA, water and methanol systems. The coarse-grained nature of the models indicates that a mean energetic representation of the system is sufficient to understand the phenomena of LCST decrease.

\end{abstract}
\maketitle

\section{Introduction} 
Cononsolvency is the phenomenon in which a polymer phase separates out in a mixture of good solvents. This behavior is exhibited by many systems such as Poly(N-isopropylacrylamide) (PNiPAM), \cite{Zhang2002, Costa2002a} Poly(N,N-diethylacrylamide) (PDEA), \cite{Maeda2002, Maeda2009, Liu2015} Poly(N,N-dimethylacrylamide), \cite{Pagonis2004, Orakdogen2006}, Polyvinylalcohol \cite{Young2002,Takahashi2003} and tertiary butyl alcohol\cite{Mochizuki2016}  in aqueous solutions with different cosolvents. Our interest lies in the cononsolvency of thermoresponsive polymer solutions which exhibit a bulk phase lower critical solution temperature (LCST) accompanied by a temperature dependent coil-to-globule transition at the single chain level. Well known examples of such polymers are PNiPAM\cite{Wang1998, Wu1998, Ono2006} and PDEA\cite{Maeda2002,Zhou2008,Yijie2010} which show LCST in water and cononsolvency with addition of cosolvents such as alcohols, dimethyl sulfoxide, and tetrahydrofuran.\cite{Yamauchi2007, Winnik1993, Bischofberger2014a}  In these systems, two interlinked phenomena can be observed with the increase in cosolvent concentration: a coil-globule-coil re-entrant transition at fixed temperature, and a decrease (increase) in the LCST in the low (high) cosolvent concentration regime.\cite{Bischofberger2014a}
\subsection{Background}
\label{sec:background}
The physical origin of cononsolvency is a very important problem in the field of polymer physics  as it finds applications in polymer processing, self assembly, drug delivery and surface modification.  The current understanding of cononsolvency in thermoresponsive polymer systems is predominantly based on the ternary system of PNiPAM, water and alcohols. Through a combination of refractive index and size measurements in PNiPAM, water and methanol mixtures, Zhang and Wu \cite{Zhang2001} proposed that cononsolvency is driven purely by the formation of water-methanol clusters. These clusters  were shown to reduce the number of hydration sites for the polymer which led to its collapse at low methanol concentrations. On the other hand, the re-entrant transition at higher methanol concentrations is driven by  solvation of the polymer by excess methanol. This mechanism is supported by the molecular simulations performed by Pang and coworkers\cite{Pang2010} on the ternary mixture of NiPAM, water, and methanol; and the DSC and FTIR measurements by Sun and Wu.\cite{Sun2010}  Bischofberger et al.\cite{Bischofberger2014,Bischofberger2014a} hypothesized, through a combination of calorimetric and scattering experiments, that the LCST is dependent on the enthalpy difference between the bulk and bound water, and the entropy loss of bound solvent. The coil-to-globule transition occurs when the entropy loss of bound solvent overcomes the energy gain of the bound solvent. The addition of alcohol reduces the energy of the bulk solvent due to kosmotropic effect which leads to the reduction in the LCST. They further observed that the  type of alcohol does not have any effect on the nature of transition. Based on these experimental observations, they proposed that a mean-field description of the solvent mixture is sufficient to explain the cononsolvency behavior in thermoresponsive polymers. The idea of mean-field description of the solvent mixture was also proposed earlier by Amiya and coworkers,\cite{Amiya1987} where they hypothesized that the attraction between the solvent(s) and cosolvent(c) or  a negative Flory-Huggins interaction parameter ($\chi_{\rm cs}$) is at the origin of cononsolvency. Given the positive values of $\chi_{\rm cs}$ in pure water-alcohol mixtures, they proposed that the change  to its negative values is due to polymer mediated interactions. 

The enhancement of the solvent-cosolvent attraction by the polymer has been questioned by Schild and coworkers\cite{Schild1991} as the behavior of PNiPAM in water-methanol mixtures remains same even with a 200 fold increase in the polymer concentration. Based on this observation, they stated that a mechanism involving local solvent-polymer interactions which vary with composition may be driving the cononsolvency behavior. Through their calculation of Kirkwood-Buff (KB) integrals for the NiPAM-water-methanol mixture, Mukherji et al.\cite{Mukherji2013, Mukherji2014} showed that the NiPAM-methanol interaction is the most dominant. They proposed that  cononsolvency is a generic phenomenon which is driven only by the preferential adsorption of the cosolvent on the polymer.\cite{Mukherji2014, Mukherji2016} They further stated that this observation combined with the argument made by Schild and coworkers\cite{Schild1991} regarding the positive nature of $\chi_{\rm cs}$ points to the non-applicability of mean-field theory. This mechanism of preferential adsorption of the cosolvent is also supported by the work of Tanaka and coworkers.\cite{Tanaka2008, Tanaka2011} They extended their cooperative hydration model\cite{Okada2005} to a two-component solvent and proposed that the competitive hydrogen bonding of the solvents with the polymer is the driving force behind cononsolvency.

However, recent studies which involve scattering experiments in combination with random phase approximation theory have shown that the effective $\chi_{\rm cs}$ is negative in  PDEA-water(s)-ethanol(c)\cite{Jia2016} and PDEA-water(s)-trimethylamine $N$-oxide (TMAO)(c) mixtures.\cite{Jia2017} This indicates that there may be polymer mediated (direct or indirect) contribution to the solvent-cosolvent interaction. Additionally, concerns have been raised by Pica and Graziano,\cite{Pica2016} and Vegt et al.\cite{Vegt2017}  regarding the usage of NiPAM monomer instead of the  polymer for calculation of KB integrals\cite{Mukherji2013, Mukherji2014} as it neglects the effects of the polymer conformational entropy on the effective coarse-grained potentials. Recent studies have also shown that preferential attraction of the cosolvent is not a prerequisite for cononsolvency.\cite{Budkov2014, Wang2017} Further, cononsolvency is absent in the case of PDEA, water and methanol mixtures even when the preferential attraction of PDEA with methanol  is higher in comparison to PNiPAM.\cite{Scherzinger2010}  Thus, it can be seen that the questions regarding the applicability of a mean-field theory are not yet  resolved. Dudowicz and coworkers \cite{Dudowicz2015, Dudowicz2015a} have remarked that the inclusion of strong associative interactions between different components in addition to the van der Waals interaction in the standard Flory-Huggins theory can lead to a description of cononsolvency from a mean-field perspective. Recent quasi-elastic neutron scattering experiments  on the ternary mixture of PNiPAM, water and methanol by Kyriakos and coworkers\cite{Kyriakos2016} indicate the presence of  polymer-water, polymer-methanol and methanol-water hydrogen bonding.  They further state that the cononsolvency phenomenon might be dependent on both the solvent-cosolvent and polymer-solvent interactions.  

In addition to these studies, there have also been efforts to understand the difference between the cononsolvency behavior of PNiPAM and PDEA in different alcohols. Richtering and coworkers studied the cononsolvency  in PNiPAM and PDEA micro-gels in water-methanol mixtures through SANS measurements.\cite{Scherzinger2010, Hofmann2013} Based on their measurements,\cite{Hofmann2013} they proposed that the complexation of amide proton  with methanol drives the cononsolvency behavior. Dalgicdir and coworkers\cite{Dalgicdir2017} studied the cononsolvency of PNiPAM in water-methanol mixtures by means of molecular simulations. They proposed that methanol hinders the ability of water to form hydrogen bond with the amide proton of PNiPAM leading to the collapse of the polymer. Both of these studies point out that the lack of LCST decrease in the ternary mixture of PDEA, water and methanol is due to the absence of the amide proton.  Since, PDEA  exhibits cononsolvency in mixtures of water with higher alcohols such as ethanol and propanol, it raises questions about the role of the amide proton. \cite{Maeda2002, Maeda2009, Liu2015} 
 
\subsection{Cononsolvency in kosmotropic cosolvents} 
\label{sec:cononsolvency}
Based on the discussion in Sec.~\ref{sec:background}, it can be seen that the explanations for cononsolvency range from specific features such as preferential attraction of the cosolvent, presence of amide proton, solvent cosolvent clustering to generic features such as mean energetics of bulk solvent. The dominating mechanisms may also be different in  the low and the high concentration regimes. Moreover, some of these mechanisms are applicable only to PNiPAM, but are not be valid in the case of  other thermoresponsive polymers.  Therefore, it can be seen that even for the subset of thermoresponsive polymers in alcohols, the understanding of the cononsolvency phenomenon is incomplete. Hence, to understand the overall behavior in the family of thermoresponsive polymers, there is a need to identify the generic underlying interactions and the regime in which they are dominant.
 
The cosolvents can be divided into two types, kosmotropic (order-making) and chaotropic (order-breaking).\cite{Moelbert2004} The phenomenon of cononsolvency can be classified on the basis of two main factors, namely, concentration and cosolvent  type. From the concentration point of view, low (high) concentration is the regime where the LCST decreases (increases) with increase in cosolvent concentration. In this work, we use coarse-grained simulations and theoretical models based on the Kolomeisky-Widom potential for studying the effects of cosolvents on the LCST. Our focus lies on kosmotropic  cosolvents in the low concentration limit. These are liquids which decrease the energy of bulk water by strengthening the hydrogen bonded network of water.\cite{Lama1965} Furthermore, these cosolvents prefer to stay in the bulk.\cite{Moelbert2004} Some of the kosmotropic cosolvents which exhibit a decrease in the LCST in aqueous solution of PNiPAM and PDEA are methanol, ethanol, propanol and TMAO.\cite{Bischofberger2014,Schroer2016,Jia2017} Our results show that the decrease in enthalpy of the bulk solvent mixture due to addition of cosolvent is responsible for the decrease in the LCST. In addition, we propose that a larger decline in the enthalpy of bulk solvent mixture is required with increase in hydrophilicity of the polymer to observe a change in LCST.  Further, from a phenomenological point of view, a mean-field description of the solvent  and solvent-cosolvent interaction is sufficient to understand the LCST decrease.
The rest of the paper is organized as follows: in Sec.~\ref{sec:models}, we introduce the simulation and theoretical models. Section~\ref{sec:results} presents the simulation results and the numerical calculations of the theoretical models. Our findings are summarized in Sec.~\ref{sec:sum}.

\section{Models}
\label{sec:models}
 In this section, we discuss  the models used in the simulations and theoretical approaches. The simulations were carried out by using a bead-spring polymer chain in solvent-cosolvent mixture. For the theoretical studies, a three-component Flory-Huggins mean-field theory with two-body interactions was employed. The scope of these models are limited to  kosmotropic (order-making) cosolvents in the low concentration limit.  The simulation and theoretical models will be explained in detail below. Unless otherwise mentioned, the term cosolvent will refer to kosmotropic cosolvents.
 
\subsection{Coarse-grained explicit solvent simulations}

Following our earlier study on the LCST of polymer in pure solvent,\cite{Bharadwaj2017}  the polymer  is modeled as a linear chain consisting of alternating solvophobic and amphiphilic beads ($N$ total beads, $N/2$ solvophobic, and $N/2$ amphiphilic beads).  To avoid any temperature dependence on the interaction potentials, the solvent-cosolvent mixture has been included explicitly by $N_{\rm s}$ solvent and $N_{\rm c}$ cosolvent beads. The potential energy for the system is given by the following expression
 \begin{eqnarray}\label{eq:energy_simu}
 E=\sum_{i=1}^{N-1} k_{\rm b}(b_{i}-b_{i0})^{2} + \sum_{i=1}^{N_{\rm t}}\sum_{j>i} 4\epsilon_{ij}\Bigg[\left(\frac{\sigma}{r_{ij}}\right)^{12}-\left(\frac{\sigma}{r_{ij}}\right)^{6} - \left(\frac{\sigma}{r_{{\rm c},ij}}\right)^{12}+\left(\frac{\sigma}{r_{{\rm c}, ij}}\right)^{6}\Bigg],
 \end{eqnarray}
where $N$ is the number of  beads in the polymer chain, as mentioned before, $k_{\rm b}$ the force constant for the bonded interaction, $b_{i}$ the bond length between neighboring beads, $N_{\rm t}$  the total number of beads in the system ($N + N_{\rm s} + N_{\rm c}$), $b_{i0}$ the equilibrium bond length and $r_{ij}$ the distance between two non-bonded beads. The second term is the Shifted Lennard-Jones (SLJ)  potential with $r_{{\rm c},ij}$ being the cutoff distance at which the potential is truncated and shifted to zero. The above form of SLJ potential ensures that all the beads are spherically symmetric  and have size  $\sigma$. All the interaction parameters are kept independent of the temperature. We define dimensionless quantities as  $\overline{r}_{ij}=r_{ij}/\sigma$, $\overline{\epsilon}_{ij}=\epsilon_{ij}/\epsilon_{\rm \textsc{ss}}$, $\overline{k}_{\rm b}=\sigma^{2}k_{\rm b}/\epsilon_{\rm \textsc{ss}}$, $\overline{b}_{i0}=b_{i0}/\sigma$, $\overline{T}=k_{\rm B}T/\epsilon_{\rm \textsc{ss}}$ and $\overline{t}= t\sqrt{\epsilon_{\rm \textsc{ss}}/(m\sigma^2)}$, where $\epsilon_{\rm \textsc{ss}}$ is the potential energy of interaction between two solvent beads. We fix the values to $\overline{b}_{i0}=1$ and $\overline{k}_{\rm b}=200$ for all the simulations. Kosmotropic cosolvents do not affect the properties of the hydration shell of the polymer as they prefer to stay in the bulk. To incorporate this feature, the interaction of the polymer with the solvent and the cosolvent is kept the same. In other words, the polymer does not distinguish between the cosolvent and solvent molecules. The values of the fixed interaction parameters are given in Table~\ref{table:1}. The parameters $\overline{\epsilon}_{\rm \textsc{as}}$, $\overline{\epsilon}_{\rm \textsc{ah}}$ and $\overline{\epsilon}_{\rm \textsc{aa}}$ are the interaction energies of the amphiphilic bead with the solvent bead, the solvophobic bead, and the amphiphilic bead, respectively, and $\overline{\epsilon}_{\rm \textsc{cs}}$ is the interaction energy of the  solvent with the cosolvent. The variation in $\overline{\epsilon}_{\rm \textsc{cs}}$ is equivalent to changing the cosolvent. We define the fraction of the cosolvent in the solvent-cosolvent mixture as $X_{\rm c}=N_{\rm c}/(N_{\rm s}+N_{\rm c})$. 
\begin{widetext}
 \begin{table}[h]
\caption{Interaction parameters of the SLJ potential. Amphiphilic, solvophobic, solvent and cosolvent are represented by A, H, S and C, respectively.}
\begin{center}
 \begin{tabular}{c c c c c c c c c}
 \hline
$ij$ \ \ \ \ & AA\ \ \ \ &HH\ \ \ \ &SS(CC)\ \ \ \ &AH\ \ \ \ &HS(HC)\ \ \ \ &AS(AC)\ \ \ \ &CS\\
 \hline
 $\overline{\epsilon}_{ij}\ \ \ \ \ $&1\ \ \ \ &1\ \ \ \ &1\ \ \ \ &1\ \ \ \ &1&1.7, 1.8, 2.0\ \ \ \ &1.1, 1.2, 1.3, 1.4\\
$\overline{r}_{{\rm c},ij}\ \ \ \ \ $&2.5\ \ \ \ &2.5\ \ \ \ &2.5\ \ \ \ &2.5 &$2^{1/6}$&2.5&2.5\\
\hline
\end{tabular}
\end{center}
\label{table:1}
 \end{table}
 \end{widetext}

Molecular dynamic simulations were performed in an NPT ensemble using the Nose-Hoover thermostat for different temperatures at a constant pressure $\overline{P}=\sigma^{3}P/\epsilon_{\rm \textsc{ss}}=0.002$. The trajectories were generated using the Velocity-Verlet algorithm with a time-step $\Delta \overline{t}=\Delta t\sqrt{\epsilon_{\rm \textsc{ss}}/(m\sigma^2)}=0.004$. All the studies were performed on a $N=200$ chain in 5000 ($N_{\rm s}+N_{\rm c}$) beads of the solvent mixture.  In our previous work on a polymer in a pure solvent,\cite{Bharadwaj2017} we observed a decrease in $\overline{R}_{\rm g}$ with temperature for $\overline{\epsilon}_{\rm \textsc{as}}=1.7,1.8$. Hence, the simulations were performed on three different polymer-solvent interactions; $\overline{\epsilon}_{\rm \textsc{as}}=\overline{\epsilon}_{\rm \textsc{ac}}=1.7, 1.8, 2.0$ at three different temperatures ($\overline{T}=0.55, 0.65, 0.75$). For each of these values, the effect of different cosolvents were studied by varying $\overline{\epsilon}_{\rm \textsc{cs}}$ and $X_{\rm c}$. Each system was equilibrated for $1 \times10^{8}$ steps, and the data was sampled  after every $2 \times 10^{6}$ steps. Four different initial configurations were used for averaging. All simulations were performed using open source molecular dynamics code LAMMPS.\cite{Plimpton1995}

 The simulation data were used for the calculation of different structural quantities. We calculated the radius of gyration, $R_{\rm g}$, of the polymer to monitor  the swelling of the polymer chain. The average $R_{\rm g}$ and error bars were calculated from the distribution obtained by sampling 1600 simulation replicas.
We define a dimensionless radius of gyration $\overline{R}_{\rm g}=R_{\rm g}/\sigma$ given by the following expression
\begin{equation}\label{eq:rg}
\overline{R}_{\rm g}=\sqrt{\frac{1}{N}\sum_{i=1}^{N}(\overline{r}_{ i}-\overline{r}_{\rm  cm})^{2}},
\end{equation}
where  $N=200$, $\overline{r}_{\rm cm}$ and $\overline{r}_{i}$ are the dimensionless coordinates of the centre of mass of the polymer chain and the $i$-th bead, respectively.

The effective interaction between the polymer beads, $U_{\rm AH}$, was calculated from the  radial distribution function of  the amphiphilic and solvophobic bead pairs  using the following expression:\cite{Chandler1987}
\begin{equation}\label{eq:pmf}
\frac{\overline{U}_{\rm AH}(\overline{r})}{\overline{T}}=-\ln{g_{\rm AH}(\overline{r})}.
\end{equation}
where $\overline{U}_{\rm AH}=U_{\rm AH}/\epsilon_{\rm ss}$ is the dimensionless effective interaction.

\subsection{Theoretical model}
\label{sec:theorymodel}
\begin{figure}
\begin{center}    
\includegraphics[scale=0.35]{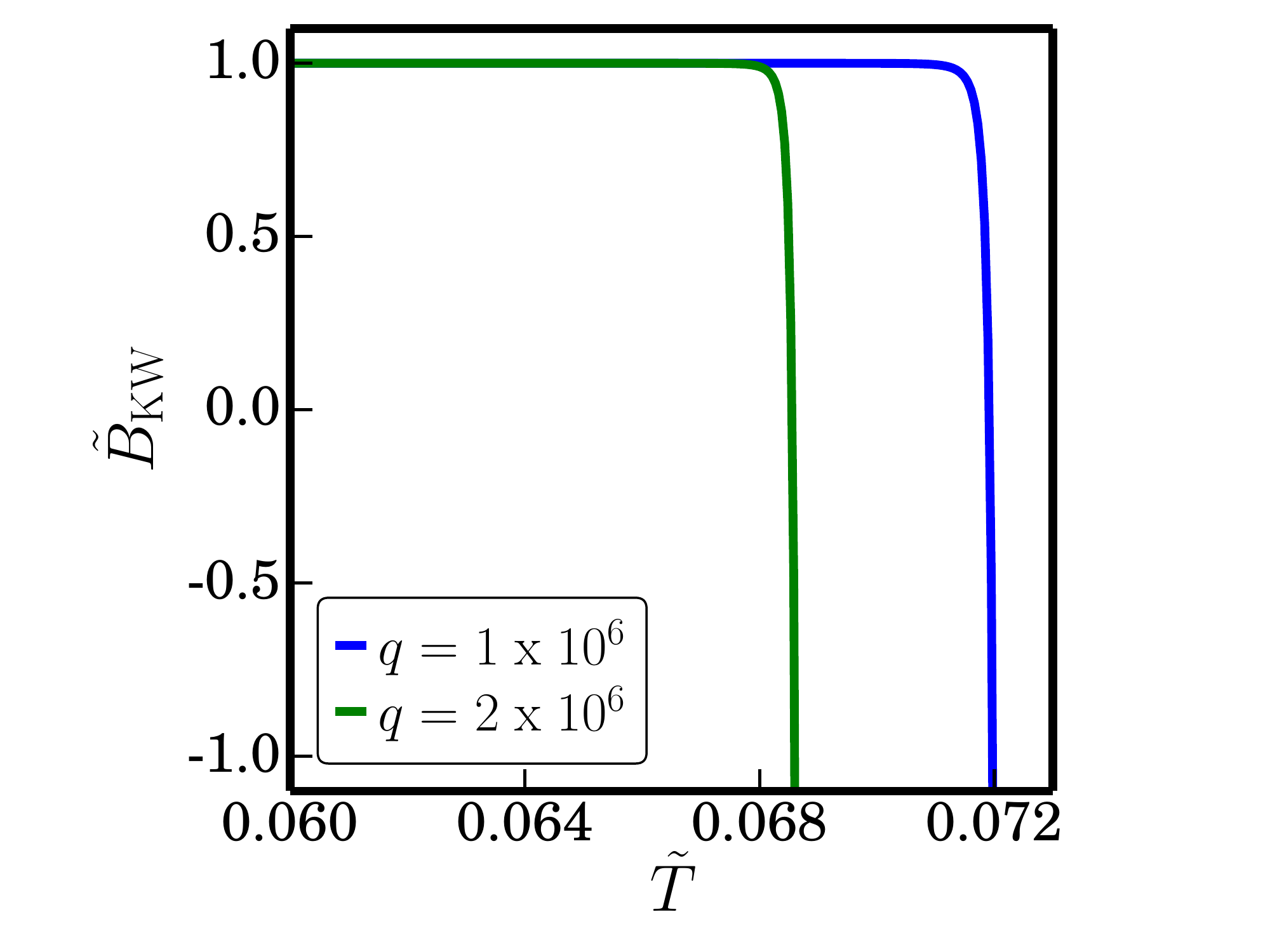}
\end{center}
\caption{Variation of $\tilde{B}_{\rm KW}$ with $\tilde{T}=k_{\rm B}T/(u-w)$ for different $q$ values.\cite{Bharadwaj2017}}
\label{fig:br}
\end{figure}

 In this section, we discuss  our  theoretical models used for studying polymer (p) in a solvent cosolvent mixture. The solvent(s)-cosolvent(c) mixture have been modeled  by the Flory-Huggins mean-field theory. The interactions between the different components (${\rm p,c,s}$) have been taken into account through the Flory-Huggins interaction parameters ($\chi_{\alpha \beta}$).  The polymer-solvent mixture exhibits an LCST and the corresponding interaction parameter $\chi_{\rm ps}$ takes into account this behavior. The LCST behavior has been modeled by using the Kolomeisky-Widom (KW) potential.\cite{Kolomeisky1999, Bharadwaj2017}  In the KW-model, solvent molecules form a one-dimensional lattice with a nearest neighbor interaction and each solvent molecule can exist in $q$ different  states. Depending on the interaction between the neighboring solvent molecules, they can be classified into one Bound State (BS) and  $q-1$ Unbound States (US) with energy $w$ and $u$, respectively. The energy of the BS state is lower than that of  the US state, $w<u$. On the other hand, the US state has higher entropy than the BS state, $k_{\rm B}\ln{q-1}>0$. The polymer-solvent interaction in the present model and the KW potential are related by the following relation (see SI for details),
 \begin{equation}\label{eq:chitob}
\chi_{\rm ps}=\frac{1}{2}-\frac{B_{\rm KW}}{\nu},
 \end{equation} 
 where $B_{\rm KW}$ is the second virial coefficient corresponding to the KW potential, $T$ is the temperature and $\nu$ is the volume per unit site. The expression for $B_{\rm KW}$ has been derived in our previous work (see SI for details).\cite{Bharadwaj2017} We define dimensionless quantities, $\tilde{B}_{\rm KW}=B_{\rm KW}/\nu$ and $\tilde{T}=k_{\rm B}T/(u-w)$. The dependence of $\tilde{B}_{\rm KW}$ on $\tilde{T}$ is shown in Fig.~\ref{fig:br} which shows that the solute-solute interaction is repulsive at low $\tilde{T}$ and attractive at high $\tilde{T}$.

Kosmotropic cosolvents prefer to stay in the bulk. This is taken into account by fixing the polymer-cosolvent interaction ($\chi_{\rm pc}$) to zero. The solvent-cosolvent Flory-Huggins  interaction parameter $\chi_{\rm cs}$ can be divided into contributions from the  enthalpic and entropic contributions. 
 \begin{equation}\label{eq:conformation}
    \chi_{\rm cs}=(\chi_{\rm cs})_{\rm enthalpic}+(\chi_{\rm cs})_{\rm entropic}.
 \end{equation}
 The entropic contribution arises from the conformational changes of the solvent on mixing. In  the case of pure water-alcohol mixtures, $(\chi_{\rm cs})_{\rm enthalpic}$ is negative and $(\chi_{\rm cs})_{\rm entropic}$ is positive. In the presence of a long chain polymer, it is reasonable to assume that the polymer induced changes to the conformational entropy of the solvent are more dominant than the cosolvent induced changes. Hence, in the case of a polymer solution, one can neglect $(\chi_{\rm cs})_{\rm entropic}$ which in turn keeps  $\chi_{\rm cs}$ to be negative. This is supported by  recent studies involving scattering experiments in combination with random phase approximation theory which show that the effective $\chi_{\rm cs}$ is negative in  PDEA-water(s)-ethanol(c)\cite{Jia2016} and PDEA-water(s)-TMAO(c) mixtures.\cite{Jia2017} The domination of the enthalpic contribution of the cosolvent over its entropic contribution in the presence of the polymer has also been suggested by Bischofberger et al.\cite{Bischofberger2014} Therefore, Eq.~(\ref{eq:conformation}) can be modified to the following form in the case of a low concentration regime
 \begin{equation}\label{eq:chics}
 \chi_{\rm cs}\approx \left(\chi_{\rm cs}\right)_{\rm enthalpic} =\frac{\Delta H_{\rm E}}{X_{\rm c}(1-X_{\rm c})}\approx \frac{\Delta H_{\rm E}}{X_{\rm c}},
 \end{equation}
where $X_{\rm c}=\phi_{\rm c}/\left(\phi_{\rm s}+\phi_{\rm c}\right)$ with  $\phi_{\rm c}$ and $\phi_{\rm s}$ being the volume fractions of the cosolvent and solvent respectively, and $\Delta H_{\rm E}$  is the enthalpy of mixing between the solvent and cosolvent. The  theoretical framework for the multiple polymer chain and single polymer chain systems are explained below.
\label{sec:model}

\subsubsection{Multiple chain solution}
\label{sec:mchain}
To study the phase separation in the bulk polymer solution, we adopt the three-component Flory-Huggins theory with only two-body interactions. The free energy of the system is given by
\begin{equation}\label{eq:mchain}
f=\frac{F}{k_{\rm B}T}=\frac{\phi_{\rm p}}{N}\ln{\phi_{\rm p}}+\phi_{\rm c}\ln{\phi_{\rm c}}+\phi_{\rm s}\ln{\phi_{\rm s}}+\chi_{\rm cs}\phi_{\rm c}\phi_{\rm s}+\chi_{\rm ps}\phi_{\rm p}\phi_{\rm s}+\chi_{\rm pc}\phi_{\rm p}\phi_{\rm c},
\end{equation}
where $F$ is the free energy per unit volume, $N$ is the degree of polymerization of the polymer chain, $\phi_{\rm p}$, is  the volume fraction of the polymer. The first three terms in the above expression are the entropic contributions, and the last three terms are the enthalpic contributions from the two-body interactions. The three volume fractions are not independent as the overall number of sites is fixed,
\begin{equation}
\phi_{\rm c}+\phi_{\rm s}+\phi_{\rm p}=1.
\end{equation} 
Hence, the free energy expression in Eq.~(\ref{eq:mchain}) takes the following form,
\begin{widetext}
\begin{equation}\label{eq:multichain}
\begin{split}
f&=\frac{\phi_{\rm p}}{N}\ln{\phi_{\rm p}}+\phi_{\rm c}\ln{\phi_{\rm c}}+(1-\phi_{\rm p}-\phi_{\rm c})\ln{(1-\phi_{\rm p}-\phi_{\rm c})}+\chi_{\rm cs}\phi_{\rm c}(1-\phi_{\rm p}-\phi_{\rm c})\\&+\chi_{\rm ps}\phi_{\rm p}(1-\phi_{\rm p}-\phi_{\rm c})+\chi_{\rm pc}\phi_{\rm p}\phi_{\rm c}.
\end{split}
\end{equation}
\end{widetext}
The phase behavior of the system is characterized by the spinodal obtained by
 \begin{equation}\label{eq:determinant}
D=\frac{\partial^{2} f}{\partial \phi_{\rm c}^{2}}\frac{\partial^{2} f}{\partial \phi_{\rm p}^{2}}-\left(\frac{\partial^{2} f}{\partial \phi_{\rm p}\partial \phi_{\rm c}}\right)^{2}=0.
\end{equation}
The  $\chi_{\rm ps}^{\rm spi}$ can be analytically obtained by solving Eq.~(\ref{eq:determinant}) (see SI for details). The variation of $\chi_{\rm ps}^{\rm spi}$ was studied for different cosolvent concentrations and $\chi_{\rm cs}$. The spinodal transition temperature $\tilde{T}_{\rm c}$, can be calculated from $\chi_{\rm ps}^{\rm spi}$ using Eq.~(\ref{eq:chitob}). 

\subsubsection{Single chain solution}
\label{sec:singlechain}
\begin{figure}[h]
\begin{center}    
\includegraphics[scale=0.14]{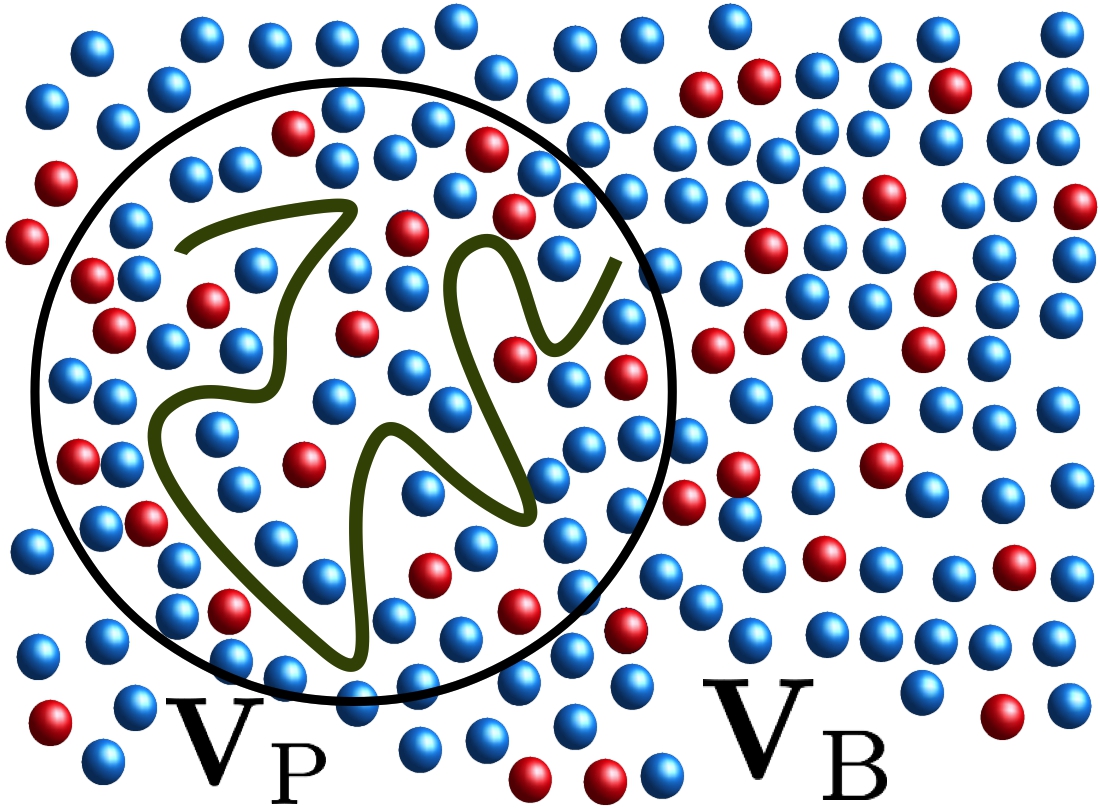}
\end{center}
\caption{Schematic of the single polymer chain in a solvent cosolvent mixture. $V_{\rm P}=4\pi R_{\rm g}^{3}/3$ is the volume occupied by the polymer, and $V_{\rm B}$ is the bulk solvent mixture.}
\label{fig:schemctog}
\end{figure}

To study the coil-to-globule transition, we consider a single polymer chain in a solvent-cosolvent mixture. The schematic of the system is given in Fig.~\ref{fig:schemctog}. Here $V_{\rm p}$ is the volume occupied by the polymer and is given by the following expression,
\begin{equation}
V_{\rm p}=\frac{4\pi R_{\rm g}^{3}}{3},
\end{equation} 

\noindent where $R_{\rm g}$ is the radius of gyration, $V_{\rm B}$ is the volume occupied by the bulk solvent-cosolvent mixture. The overall volume of the system $V(=V_{\rm P}+V_{\rm B})$ is fixed. The free energy of the system is given by the following expression,
\begin{equation}
F_{\rm cg}=F_{\rm elastic} + F_{\rm mixing},
\end{equation}
where $F_{\rm elastic}$ is the elastic contribution due to the structural changes in the polymer chain and $F_{\rm mixing}$ is the contribution arising due to the entropy of mixing and two-body interactions. The elastic free energy $F_{\rm elastic}$, has been adopted from the theory of coil-to-globule transition.\cite{Grosberg1992}

\begin{equation}
f_{\rm elastic}=\frac{F_{\rm elastic}}{k_{\rm B}T}=\alpha^{2}+\frac{1}{\alpha^{2}},
\end{equation}
where $\alpha=R_{\rm g}/R_{\rm g}^{\rm ideal}$ characterizes the extent of swelling. For $F_{\rm mix}$, we adopt the Flory-Huggins mean-field theory,
\begin{equation}
  \begin{split}
f_{\rm mixing}=\frac{F_{\rm mixing}}{k_{\rm B}T}& = \frac{V_{\rm P}}{v}(\phi_{\rm s}^{'}\ln{\phi_{\rm s}^{'}}+\phi_{\rm c}^{'}\ln{\phi_{\rm c}^{'}}+\chi_{\rm ps}\phi_{\rm p}^{'}\phi_{\rm s}^{'}+\chi_{\rm cs}\phi_{\rm c}^{'}\phi_{\rm s}^{'}+\chi_{\rm pc}\phi_{\rm p}^{'}\phi_{\rm c}^{'})\\ &+\frac{V_{\rm B}}{v}(\phi_{\rm s}^{''}\ln{\phi_{\rm s}^{''}}+\phi_{\rm c}^{''}\ln{\phi_{\rm c}^{''}}+\chi_{\rm cs}\phi_{\rm c}^{''}\phi_{\rm s}^{''}),
\end{split}
\end{equation}
where $v$ is the fixed volume per unit site, $\phi_{\rm i}^{'}=N_{\rm i}v/V_{\rm P}$ and $\phi_{\rm i}^{''}=N_{\rm i}v/V_{\rm B}$ are the local volume fractions in the polymer  and the bulk  volume, respectively. The overall free energy is as follows,  
\begin{widetext}
\begin{equation}\label{eq:sinchain}
\begin{split}
  f_{\rm cg}=\frac{F_{\rm cg}}{k_{\rm B}T}&=\alpha^{2}+\frac{1}{\alpha^{2}}+\frac{V_{\rm P}}{v}(\phi_{\rm s}^{'}\ln{\phi_{\rm s}^{'}}+\phi_{\rm c}^{'}\ln{\phi_{\rm c}^{'}}+\chi_{\rm ps}\phi_{\rm p}^{'}\phi_{\rm s}^{'}+\chi_{\rm cs}\phi_{\rm c}^{'}\phi_{\rm s}^{'}+\chi_{\rm pc}\phi_{\rm p}^{'}\phi_{\rm c}^{'})\\&+\frac{V_{\rm B}}{v}(\phi_{\rm s}^{''}\ln{\phi_{\rm s}^{''}}+\phi_{\rm c}^{''}\ln{\phi_{\rm c}^{''}}+\chi_{\rm cs}\phi_{\rm c}^{''}\phi_{\rm s}^{''}).
  \end{split}
\end{equation}
\end{widetext}
Given that the overall volume ($V$) of the system is fixed, one can define the volume fractions with respect to $V$ instead of the local volumes $V_{\rm P}$ and $V_{\rm B}$. Then the dimensionless free energy $f_{\rm cg}$ can then be  rewritten in the following way (see SI for details),
\begin{widetext}
\begin{equation}\label{eq:freenergy}
\begin{split}
\frac{f_{\rm cg}}{V/v}&=\frac{(\Phi_{\rm p}^{'})^{5/3}}{N^{2/3}(\Phi^{\rm '})^{2/3}\kappa^{2/3}} +\frac{(\Phi^{\rm '})^{2/3}(\Phi_{\rm p}^{'})^{1/3}\kappa^{2/3}}{N^{4/3}}+\Phi^{\rm '}_{\rm c}\ln{\frac{\Phi_{\rm c}^{\rm '}}{\Phi^{\rm '}}}+\Phi_{\rm s}^{\rm '}\ln{\frac{\Phi_{\rm s}^{\rm '}}{\Phi^{\rm '}}}+\chi_{\rm ps}\frac{\Phi_{\rm p}^{\rm '}\Phi_{\rm s}^{\rm '}}{\Phi^{\rm '}}\\&+\chi_{\rm cs}\frac{\Phi^{\rm '}_{\rm c}\Phi_{\rm s}^{\rm '}}{\Phi^{\rm '}}+\chi_{\rm pc}\frac{\Phi^{\rm '}_{\rm c}\Phi_{\rm p}^{'}}{\Phi^{\rm '}}+\Phi^{\rm ''}_{\rm c}\ln{\frac{\Phi^{\rm ''}_{\rm c}}{\Phi^{\rm ''}}}+\Phi^{''}_{\rm s}\ln{\frac{\Phi^{\rm ''}_{\rm s}}{\Phi^{\rm ''}}}+\chi_{\rm cs}\frac{\Phi^{\rm ''}_{\rm c}\Phi^{\rm ''}_{\rm s}}{\Phi^{\rm ''}},
\end{split}
\end{equation}
\end{widetext}
where $\Phi^{'}=V_{\rm P}/V$, $\Phi^{''}=V_{\rm B}/V$, $\Phi^{'}_{\rm i}=N^{\rm '}_{\rm i}v/V$ and $\kappa$ is the parameter which controls the rigidity of the polymer chain  and has been set to unity for all the cases. The equilibrium swelling of the polymer chain can  be obtained by minimizing the free energy in Eq.~(\ref{eq:freenergy}) with respect to $\Phi^{'}$, $\Phi^{''}$, $\Phi^{'}_{\rm s}$, $\Phi^{'}_{\rm c}$, $\Phi^{''}_{\rm s}$ and $\Phi^{''}_{\rm c}$. The minimization is subjected to the following constraints, 
\begin{equation}
\begin{split}
\Phi^{\rm '}_{\rm s}+\Phi^{\rm '}_{\rm c}+\Phi_{\rm p}^{'}&=\Phi^{\rm '},\\
\Phi^{\rm ''}_{\rm s}+\Phi^{''}_{\rm c}&=\Phi^{\rm ''},\\
\Phi^{\rm ''}_{\rm c}+\Phi^{\rm '}_{\rm c}&=\Phi_{\rm c},\\
\Phi^{\rm ''}+\Phi^{\rm '}&=1,\\
\end{split}
\end{equation}
where $\Phi_{\rm c}=(N_{\rm c}^{'}+N_{\rm c}^{''})v/V$ is the fixed overall concentration of the cosolvent.  
 As the polymer is confined to the polymer volume, we have $\Phi^{'}_{\rm p}=\Phi_{\rm p}$ and $\Phi^{''}_{\rm p}=0$, where $\Phi_{\rm p}=Nv/V$ is the overall fixed concentration of the polymer. The calculations were performed using the open source optimization package IPOPT.\cite{Wachter2006} The variation of the swelling of the polymer chain with temperature has been studied for different $\chi_{\rm cs}$ and $X_{\rm c}$. As in the case of the multiple chain theory, the temperature is calculated from $\chi_{\rm ps}$ by using Eq.~(\ref{eq:chitob}).

\section{Results and Discussion}
\label{sec:results}
\subsection{Simulation}
\begin{figure}[h]
\begin{center}    
\subfigure{\label{fig:ra}\includegraphics[scale=0.35]{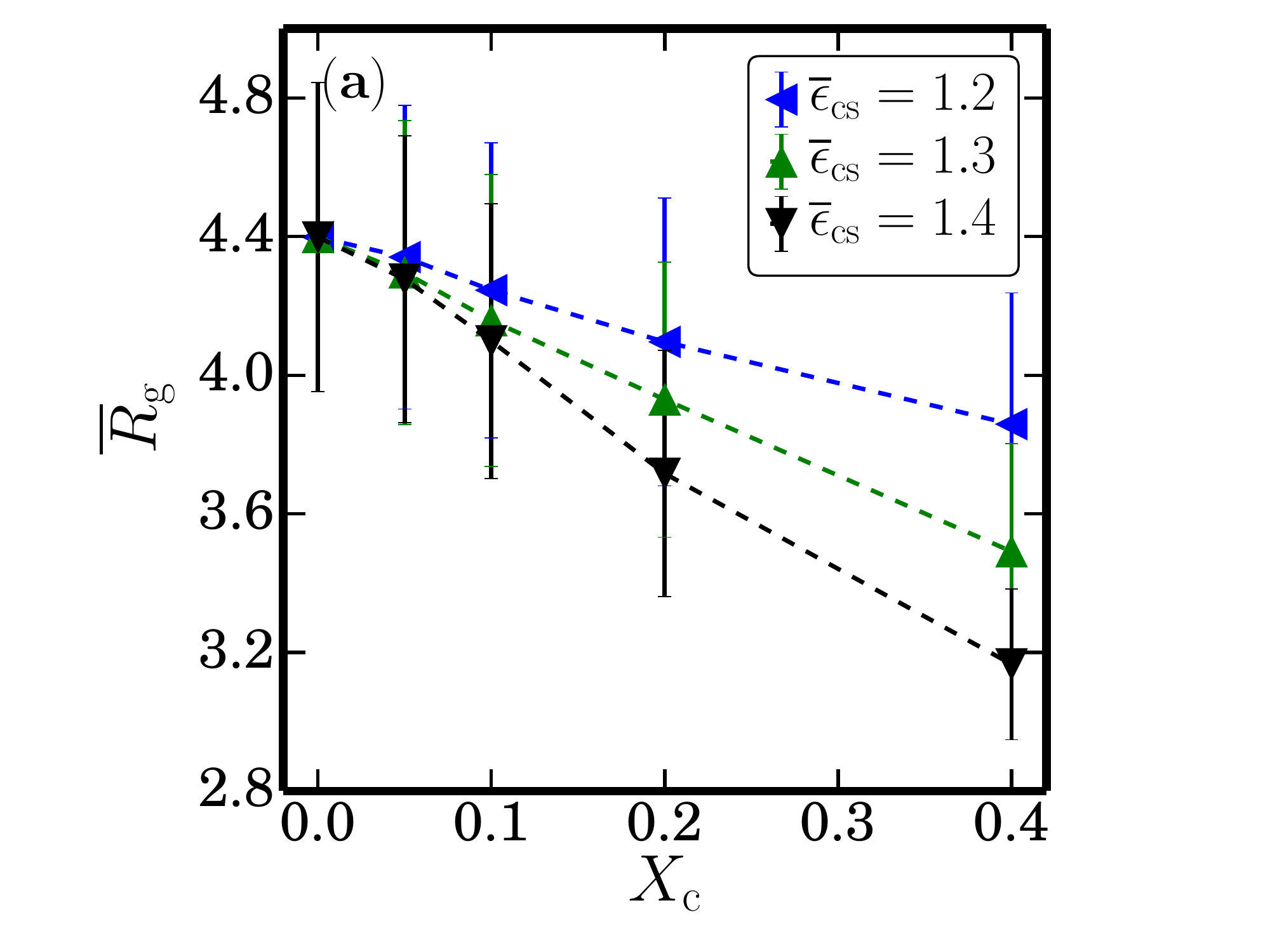}}
\subfigure{\label{fig:rb}\includegraphics[scale=0.35]{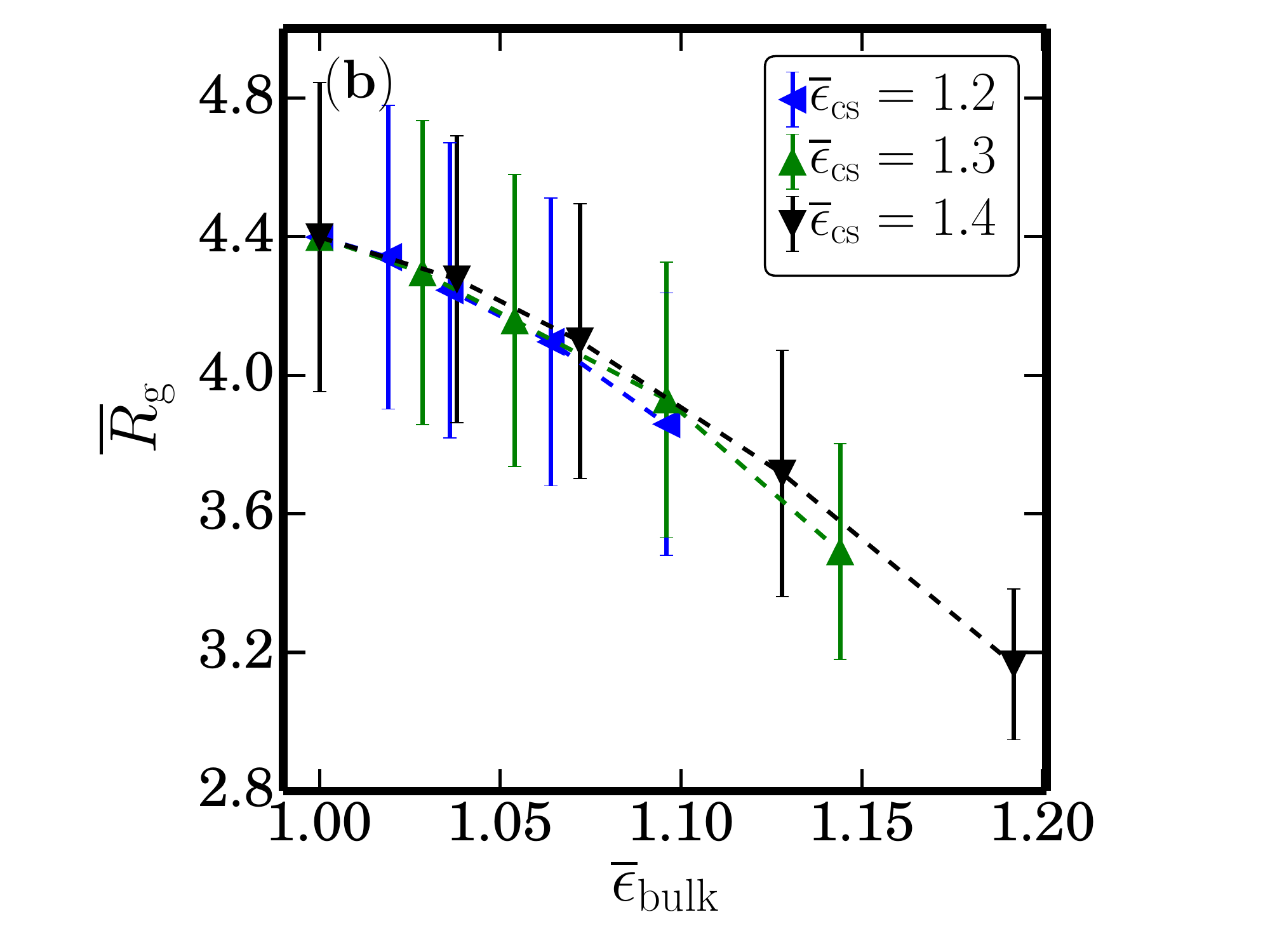}}
\end{center}
\caption{Variation of $\overline{R}_{\rm g}$ for different solvent cosolvent interaction strengths at $\overline{\epsilon}_{\rm \textsc{as}}=1.8$ and $\tilde{T}=0.65$ with (a) cosolvent fraction $X_{\rm c}$, and (b) mean energy of bulk solvent-cosolvent mixture $\overline{\epsilon}_{\rm bulk}$, which can be related to $X_{\rm c}$ by Eq.~(\ref{eq:bulk}) .}
\label{fig:rt}
\end{figure}

To understand the variation of $\overline{R}_{\rm g}$ with cosolvent concentration for different cosolvents, we first look at the system with $\overline{\epsilon}_{\rm \textsc{as}}=1.8$. In Fig.~\ref{fig:ra}, we see that for a solvent pair, increase in $X_{\rm c}$ leads to a more collapsed state, while for a given cosolvent concentration, higher interaction strength ($\overline{\epsilon}_{\rm \textsc{cs}}$)  tends to collapse the chain. Many studies\cite{Zhang2001, Bischofberger2014,Dudowicz2015, Bharadwaj2017} have shown the effective interaction of the bulk solvent is an important contribution which affects the LCST. In our study, we calculate the effective interaction of the bulk solvent-cosolvent mixture ($\overline{\epsilon}_{\rm bulk}$), which is dependent on $\overline{\epsilon}_{\rm cc}$, $\overline{\epsilon}_{\rm ss}$, $\overline{\epsilon}_{\rm cs}$ and $X_{\rm c}$, by the mixing rule commonly used in mean-field theory (see SI for details),  
\begin{equation}\label{eq:bulk}
\overline{\epsilon}_{\rm bulk}=(1-X_{\rm c})^{2}\ \overline{\epsilon}_{\rm \textsc{ss}}+2(1-X_{\rm c})X_{\rm c}\ \overline{\epsilon}_{\rm \textsc{cs}}+X_{\rm c}^{2}\ \overline{\epsilon}_{\rm \textsc{cc}}.
\end{equation}
We then perform a variable transformation from $X_{\rm c}$ to $\overline{\epsilon}_{\rm bulk}$. The variation of $\overline{R}_{\rm g}$  with $\overline{\epsilon}_{\rm bulk}$  for different $\overline{\epsilon}_{\rm cs}$ is shown in Fig.~\ref{fig:rb}. We highlight the data collapse  indicating that for a given polymer, $\overline{R}_{\rm g}$ is dependent only on the bulk solvent energy. Hence, it can be said that the mean energetic representation of the solvent-cosolvent interaction is sufficient to understand the collapse of the polymer with the addition of cosolvent. These results are in agreement with the experimental findings by Bischofberger et al.\cite{Bischofberger2014a} 

\begin{figure}[h]
\begin{center}    
\subfigure{\label{fig:raa}\includegraphics[scale=0.35]{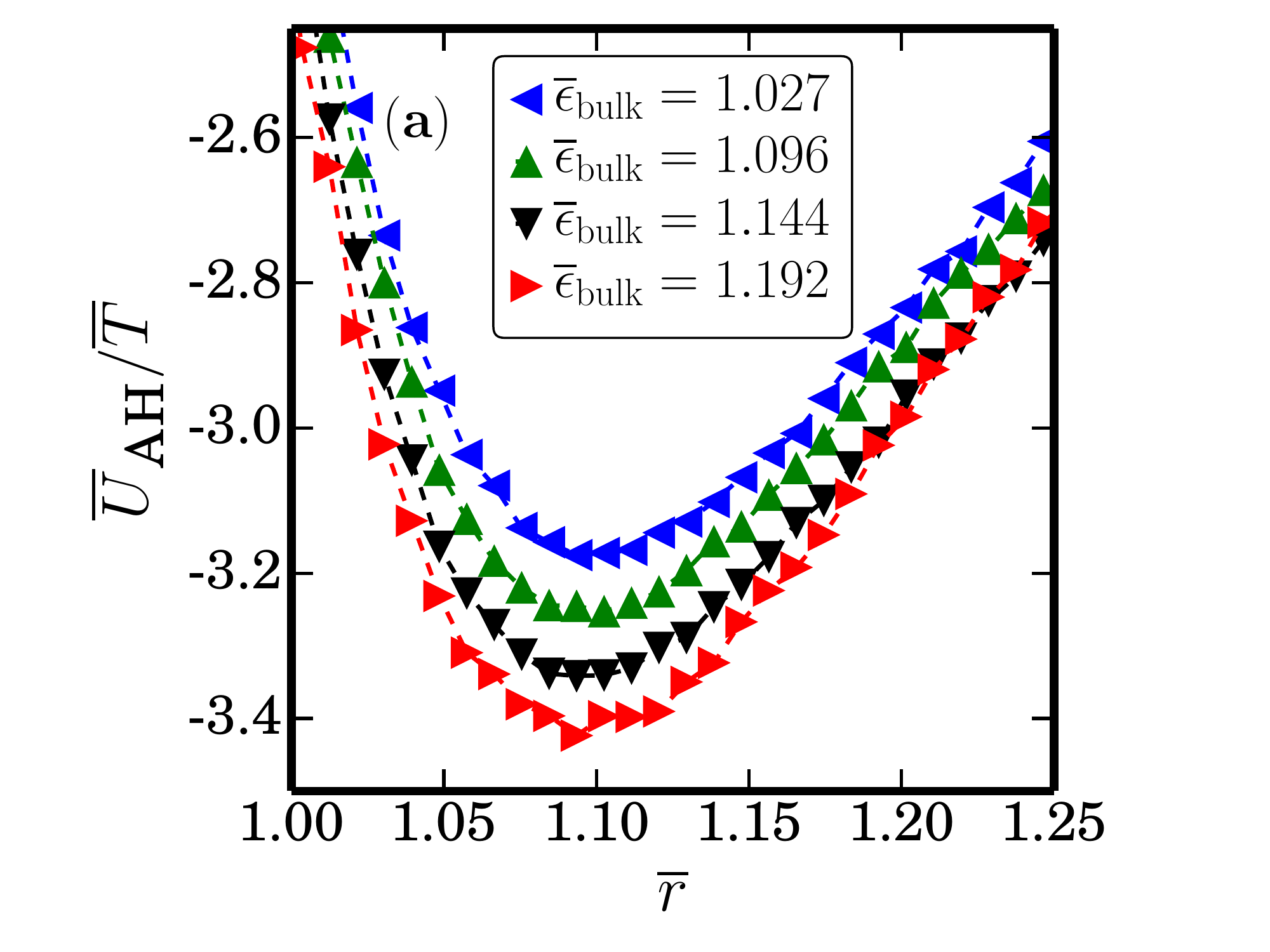}}
\subfigure{\label{fig:rbb}\includegraphics[scale=0.35]{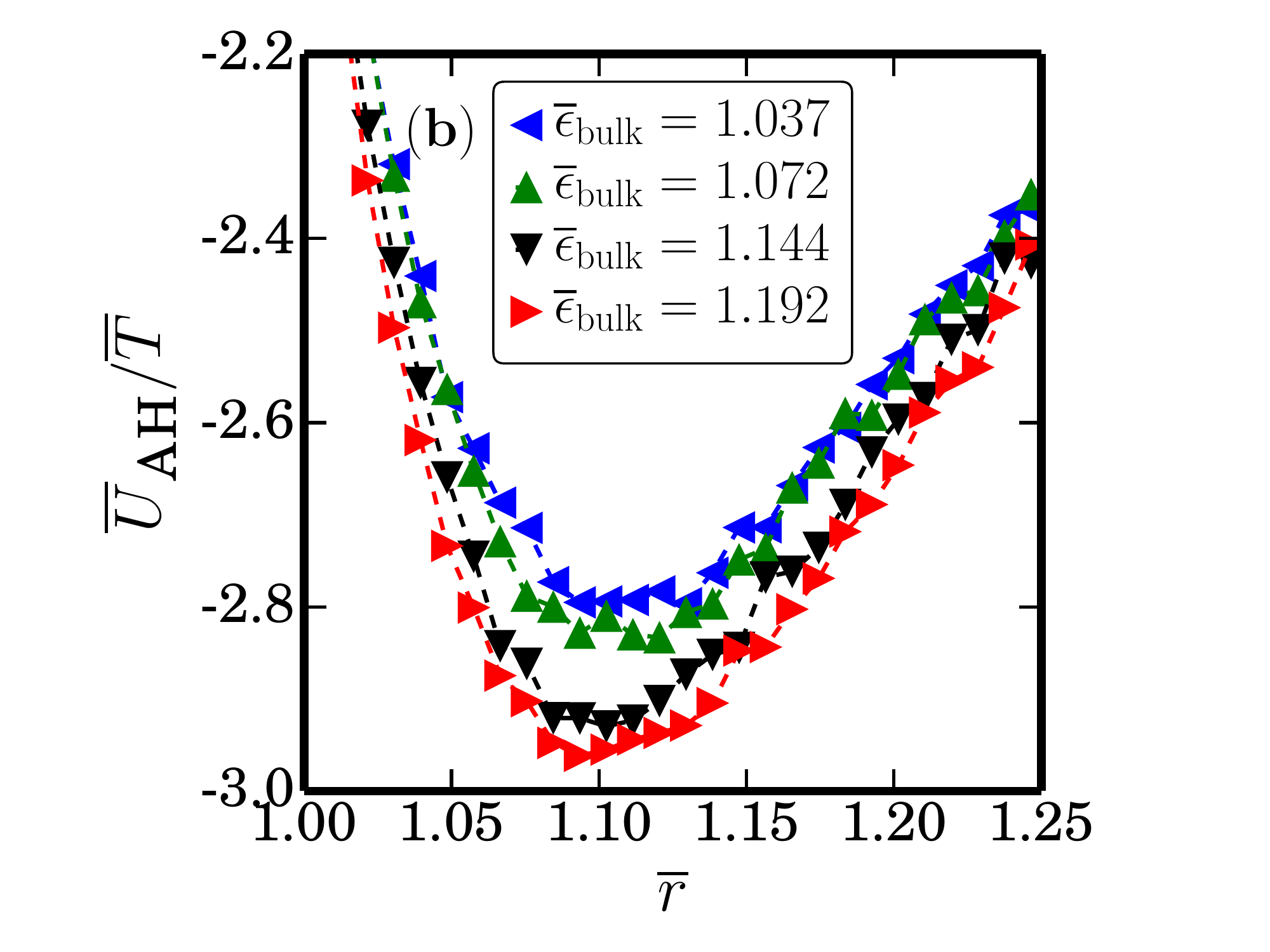}}
\end{center}
\caption{Variation of the effective attraction $\overline{U}_{\rm \textsc{ah}}/k_{\rm B}\overline{T}=-\ln{g_{\rm \textsc{ah}}}$ with $\overline{r}$ for different $\overline{\epsilon}_{\rm bulk}$ values for (a) $\overline{\epsilon}_{\rm \textsc{as}}=1.8$ and (b) $\overline{\epsilon}_{\rm \textsc{as}}=2.0$.}
\label{fig:rtt}
\end{figure}

In our earlier work on the mechanism of LCST  in pure solvent, \cite{Bharadwaj2017} we showed that the LCST is dependent on the energy difference between the bulk and bound solvent, and the entropy loss of the bound solvent. This can now be extended to the case of LCST in solvent-cosolvent mixture. In the case of pure solvent ($X_{\rm c}=0$), the polymer is an expanded state due to the domination of the mean energy difference between the bulk and bound solvent over the bound solvent entropy loss. The attraction between the beads is screened due to the high level of solvation. With the addition of the cosolvent, the energy of the bulk solvent decreases. The entropy and the energy contributions from the solvation shell of the polymer  are not modified as the cosolvent prefers to stay in the bulk. The reduction in the bulk solvent energy leads to a lower mean energy difference between the bulk and the bound solvent which causes a drop in the LCST and a decrease in  $\overline{R}_{\rm g}$. The gain in the bulk solvent energy leads to unbinding of the solvent from the solvation shell of the polymer. This leads to higher contact between the polymeric beads leading to increase in the effective  attraction between them causing the chain to collapse.  Figure~\ref{fig:rtt} shows the variation in the effective attraction, $\overline{U}_{\rm \textsc{ah}}/k_{\rm B}\overline{T}=-\ln{g_{\rm \textsc{ah}}}$, with $\overline{r}$ for different $\overline{\epsilon}_{\rm bulk}$ values. It can be seen that the attraction between the polymeric beads becomes stronger with increase in $\overline{\epsilon}_{\rm bulk}$.
The temperature dependent variation of $\overline{R}_{\rm g}$ for different cosolvent conditions is shown in Fig.~\ref{fig:rt2}. For $\overline{\epsilon}_{\rm bulk}=1.144$, $\overline{R}_{\rm g}$ is almost independent of $\overline{T}$, for the range of temperatures investigated, which implies that  the LCST is lower than $\overline{T}=0.55$. With a decrease in $\overline{\epsilon}_{\rm bulk}$, which for a given cosolvent is equivalent to drop in its concentration, $\overline{R}_{\rm g}$ increases for all the temperatures indicating a systematic rise in LCST. In other words, the LCST drops with increase in $\overline{\epsilon}_{\rm bulk}$. 
\begin{figure}[h]
\includegraphics[scale=0.35]{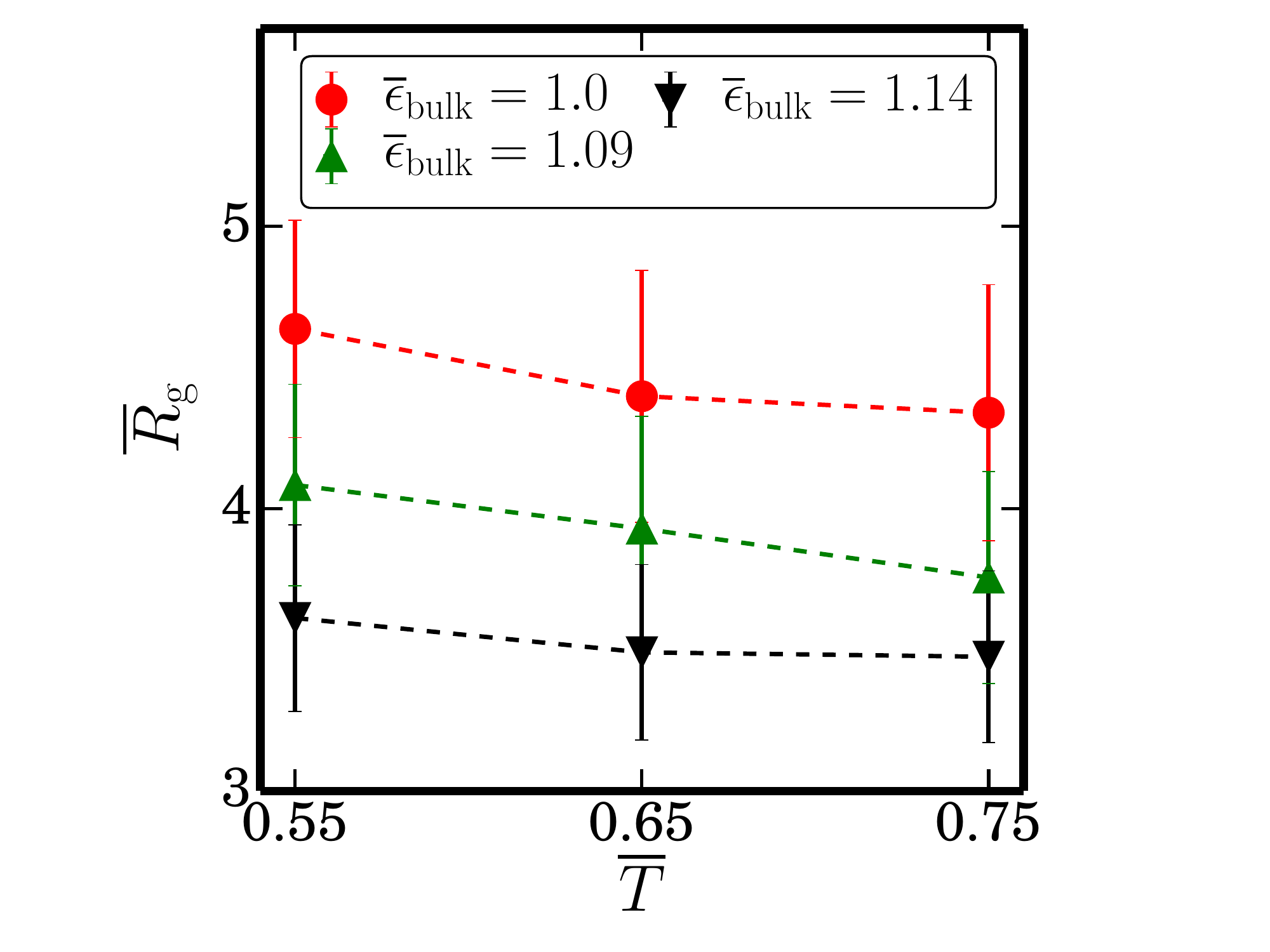}
\caption{Variation of $\overline{R}_{\rm g}$ with $\overline{T}$ for different $\overline{\epsilon}_{\rm bulk}$ values. For all the cases, $\overline{\epsilon}_{\rm \textsc{as}}=1.8$. $\overline{\epsilon}_{\rm bulk}=1.0$ is the case when $X_{\rm c}=0$.}
\label{fig:rt2}
\end{figure}

\begin{figure}[h] 
\begin{center}    
\subfigure{\label{fig:rt3}\includegraphics[scale=0.35]{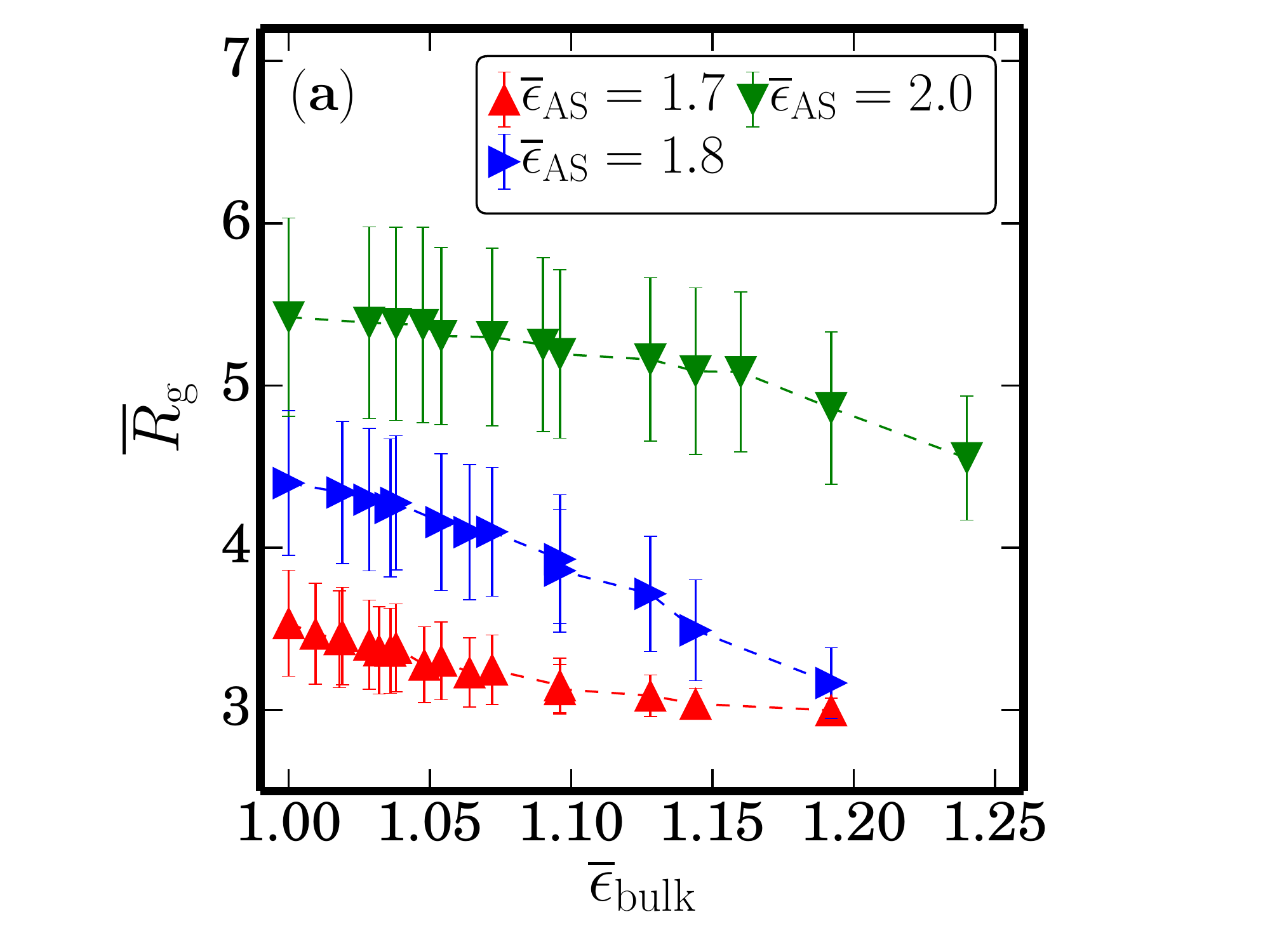}}
\subfigure{\label{fig:rt3_2}\includegraphics[scale=0.35]{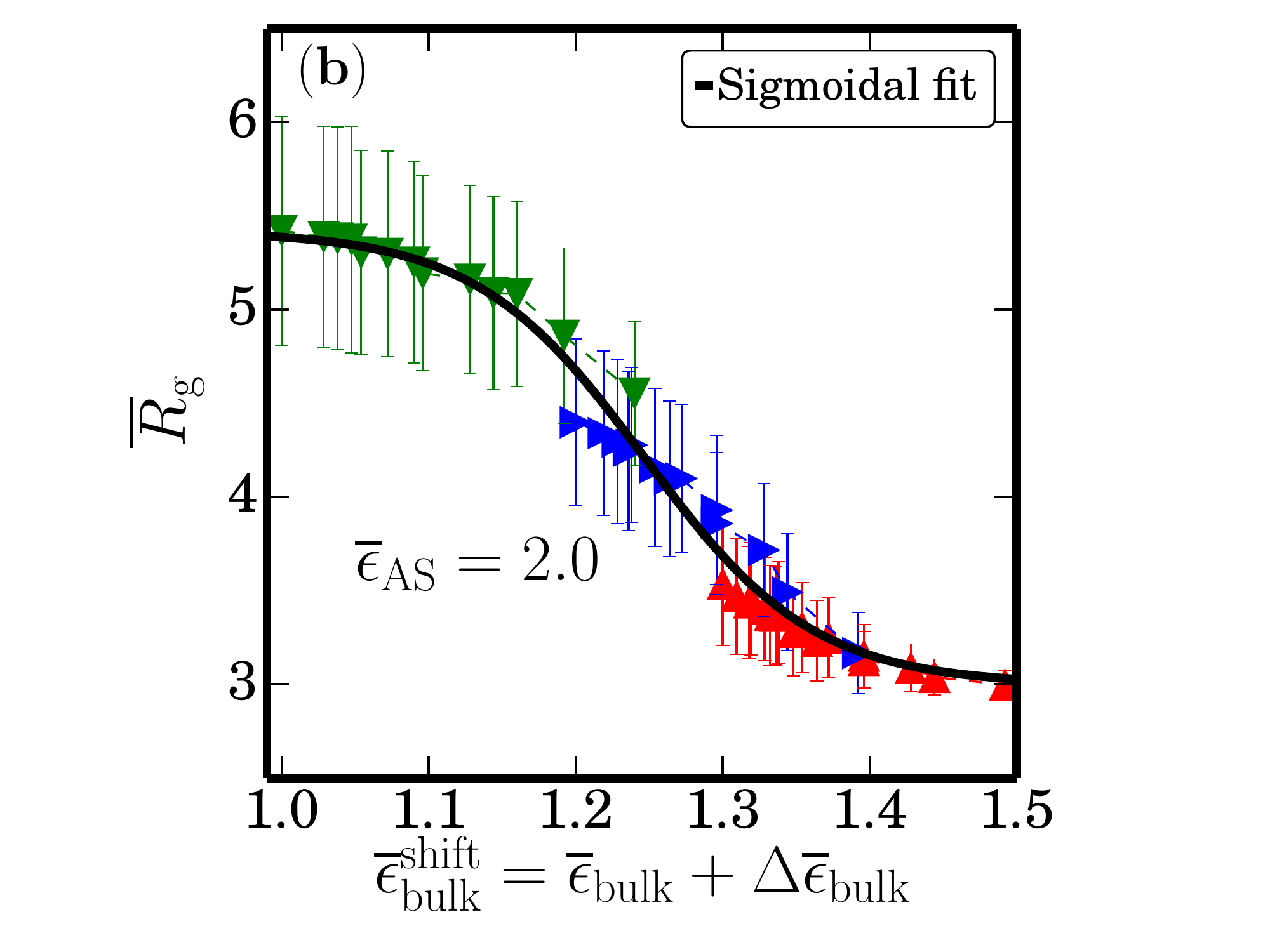}}
\end{center}
\caption{(a) Variation of $\overline{R}_{\rm g}$ with $\overline{\epsilon}_{\rm bulk}$ for different $\overline{\epsilon}_{\rm \textsc{as}}$ values, and (b) variation of  $\overline{R}_{\rm g}$ with $\overline{\epsilon}_{\rm bulk}^{\rm shift}=\overline{\epsilon}_{\rm bulk}+\Delta \overline{\epsilon}_{\rm bulk}$ for $\overline{\epsilon}_{\rm \textsc{as}}=2.0$ master curve. The sigmoidal curve has the form $\overline{R}_{\rm g}=C_{\rm 1}+a/[b+\exp{\left(c(\overline{\epsilon}_{\rm bulk}^{\rm shift}-C_{2})\right)}]$where $a=0.96, b=0.39, c=17.4, C_{1}=3.0, C_{2}=1.3$. }
\end{figure}

\begin{figure}[h]
\begin{center}
\includegraphics[scale=0.35]{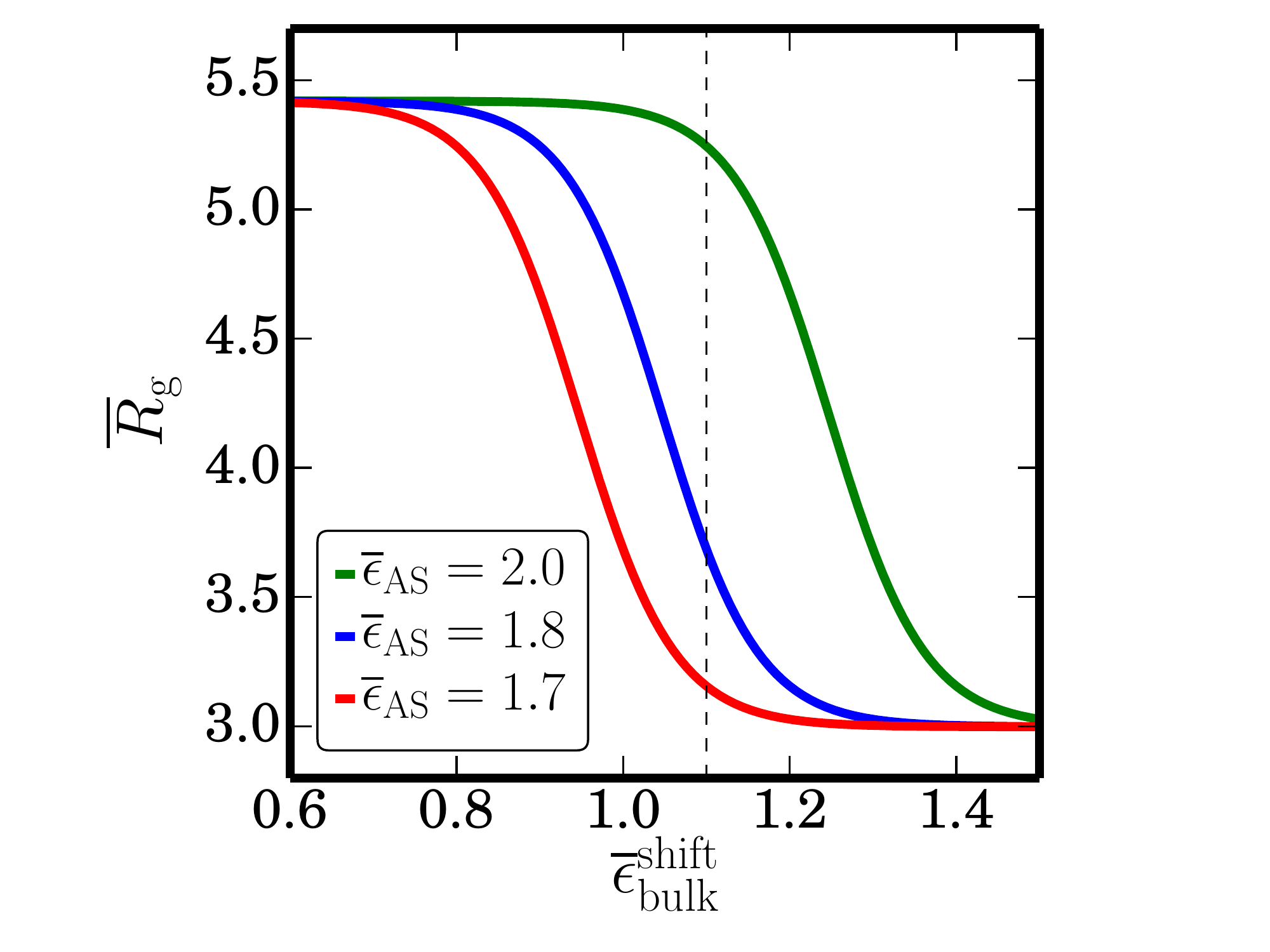}
\end{center}
\caption{Sigmoidal fits to master curves for different $\overline{\epsilon}_{\rm \textsc{as}}$ values. The curves have the form $\overline{R}_{\rm g}=C_{\rm 1}+a/[b+\exp{\left(c(\overline{\epsilon}_{\rm bulk}^{\rm shift}-C_{2})\right)}]$. See SI for more details.}
\label{fig:rt3_3}
\end{figure}

The effect of the hydrophilicity of the polymer was also studied by performing simulations for $\overline{\epsilon}_{\rm \textsc{as}}=1.7, 2.0$ with the same solvent-cosolvent combinations as before. Figure~\ref{fig:rt3} shows the variation of $\overline R_{\rm g}$ with $\overline{\epsilon}_{\rm bulk}$  
for different $\overline{\epsilon}_{\rm \textsc{as}}$ values. For a given $\overline{\epsilon}_{\rm bulk}$, it can be seen that $\overline R_{\rm g}$  increases systematically with increase in the hydrophilicity of the polymer. As the hydrophilicity of the polymer becomes higher, the energy difference between the bound and bulk solvent increases. This increases the stability of the bound solvent due to which a larger drop in the bulk solvent energy (increase in $\overline{\epsilon}_{\rm bulk}$) is required to collapse the polymer. 
Combining the effects of the polymer hydrophilicity and cosolvent induced bulk solvent energy decrease, it can be said that $\overline{R}_{\rm g}$ is dependent on the the difference between the polymer-solvent interaction energy $\overline{\epsilon}_{\rm \textsc{as}}$, and the energy of the bulk solvent-cosolvent mixture $\overline{\epsilon}_{\rm bulk}$. This implies that polymers with different hydrophilicity should have the same $\overline{R}_{\rm g}$ as long as the mean energy difference between the bound solvent and the bulk solvent-cosolvent mixture  is same. This can be used to obtain the master coil-to-globule transition curves. For example, $\overline{R}_{\rm g}$ for a system with parameters $\overline{\epsilon}_{\rm \textsc{as},1}$  and $\overline{\epsilon}_{\rm bulk,1}$ should be equivalent to $\overline{R}_{\rm g}$ for a system with parameters $\overline{\epsilon}_{\rm \textsc{as},2}$  and $\overline{\epsilon}_{\rm bulk,2}$, given that $\Delta \overline{\epsilon}_{\rm bulk}=\overline{\epsilon}_{\rm bulk,2}-\overline{\epsilon}_{\rm bulk,1}=\overline{\epsilon}_{\rm \textsc{as},2}-\overline{\epsilon}_{\rm \textsc{as},1}$. Using these relations, we shifted the trends in Fig.~\ref{fig:rt3} of polymer with $\overline{\epsilon}_{\rm \textsc{as}}=1.7$ by $\Delta \overline{\epsilon}_{\rm bulk}=0.3$ and of $\overline{\epsilon}_{\rm \textsc{as}}=1.8$ by $\Delta \overline{\epsilon}_{\rm bulk}=0.2$   to obtain the master coil-to-globule transition curve for $\overline{\epsilon}_{\rm \textsc{as}}=2.0$. The resulting master curve is shown in Figure~\ref{fig:rt3_2}. It can be seen that the data collapses well into a sigmoidal curve of the form $\overline{R}_{\rm g}=C_{\rm 1}+a/[b+\exp{\left(c(\overline{\epsilon}_{\rm bulk}^{\rm shift}-C_{2})\right)}]$ where $\overline{\epsilon}_{\rm bulk}^{\rm shift}=\overline{\epsilon}_{\rm bulk}+\Delta \overline{\epsilon}_{\rm bulk}$, indicating that $\overline{R}_{\rm g}$ is only dependent on the mean energy difference between the bound solvent and the bulk solvent-cosolvent mixture. Similar master curves have been obtained for $\overline{\epsilon}_{\rm \textsc{as}}=1.7,1.8$. The sigmoidal fits to these master curves are shown in Fig.~\ref{fig:rt3_3}.
It can be seen that a larger drop in bulk solvent-cosolvent energy (higher $\overline{\epsilon}_{\rm bulk}^{\rm shift}$) is required to collapse a polymer with higher hydrophilicity. Another point to be noted is that a cosolvent which can induce a change of bulk energy to $\overline{\epsilon}_{\rm bulk}^{\rm shift}=1.1$ is able to collapse the polymers with $\overline{\epsilon}_{\rm \textsc{as}}=1.7,1.8$ but not $\overline{\epsilon}_{\rm \textsc{as}}=2.0$. This could be a possible  reason for methanol to cause an LCST decrease in aqueous solution of PNiPAM but not in that of PDEA. This point will be discussed in detail in Sec~\ref.

In the simulation model, the solvent mixture is spherically symmetric and does not take into account the conformational entropy change of mixing. This model is able to exhibit the LCST decrease and $R_{\rm g}$ collapse which have been observed in experiments. This in turn justifies keeping only the mixing energy contribution to the  Flory-Huggins interaction parameter ($\chi_{\rm cs}$) in the theoretical model (see Eq.~(\ref{eq:chics})).
\subsection{Theoretical model}
\begin{figure}[h]
\begin{center}    
\subfigure{\label{fig:ba}\includegraphics[scale=0.35]{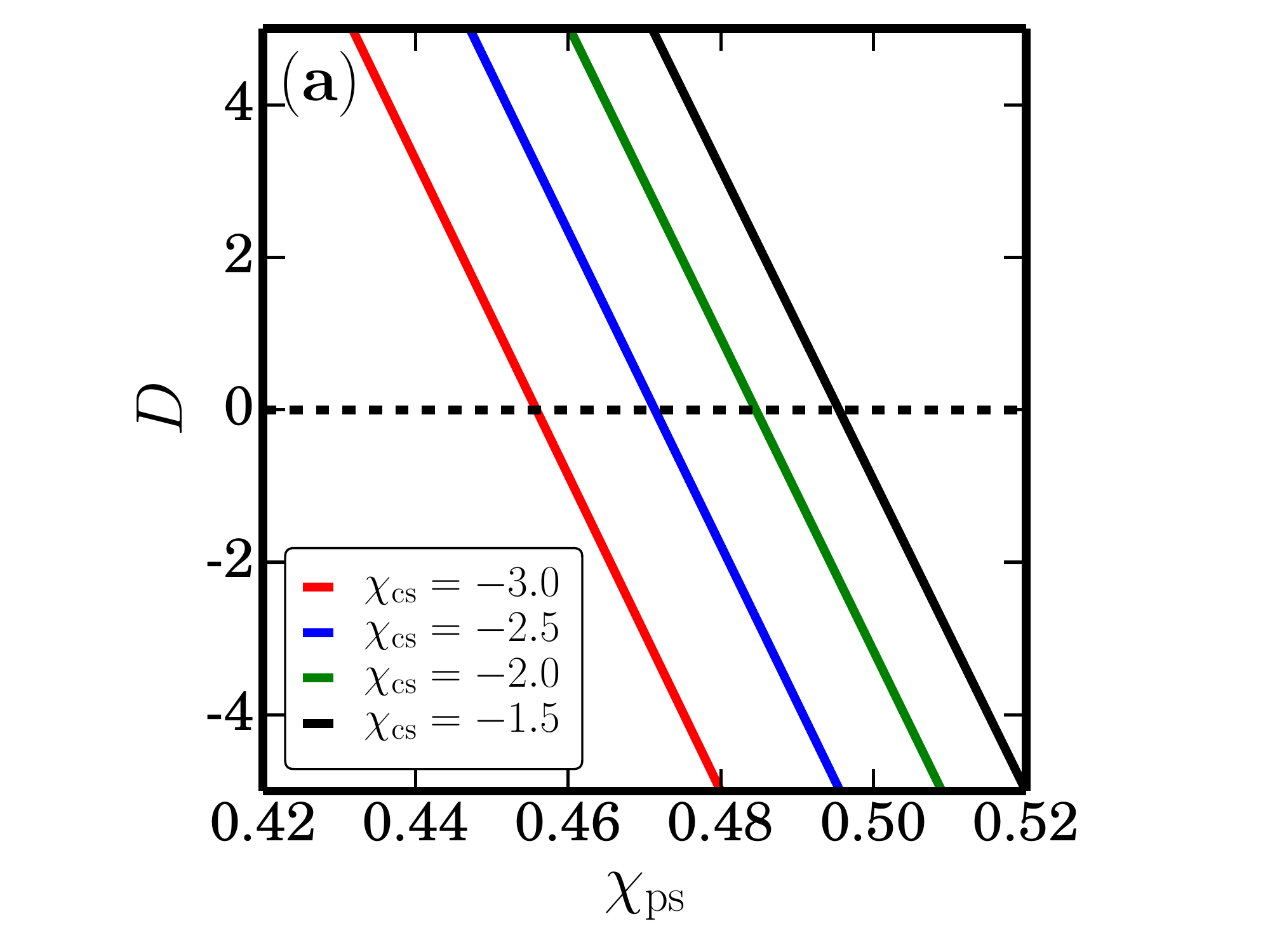}}
\subfigure{\label{fig:bb}\includegraphics[scale=0.35]{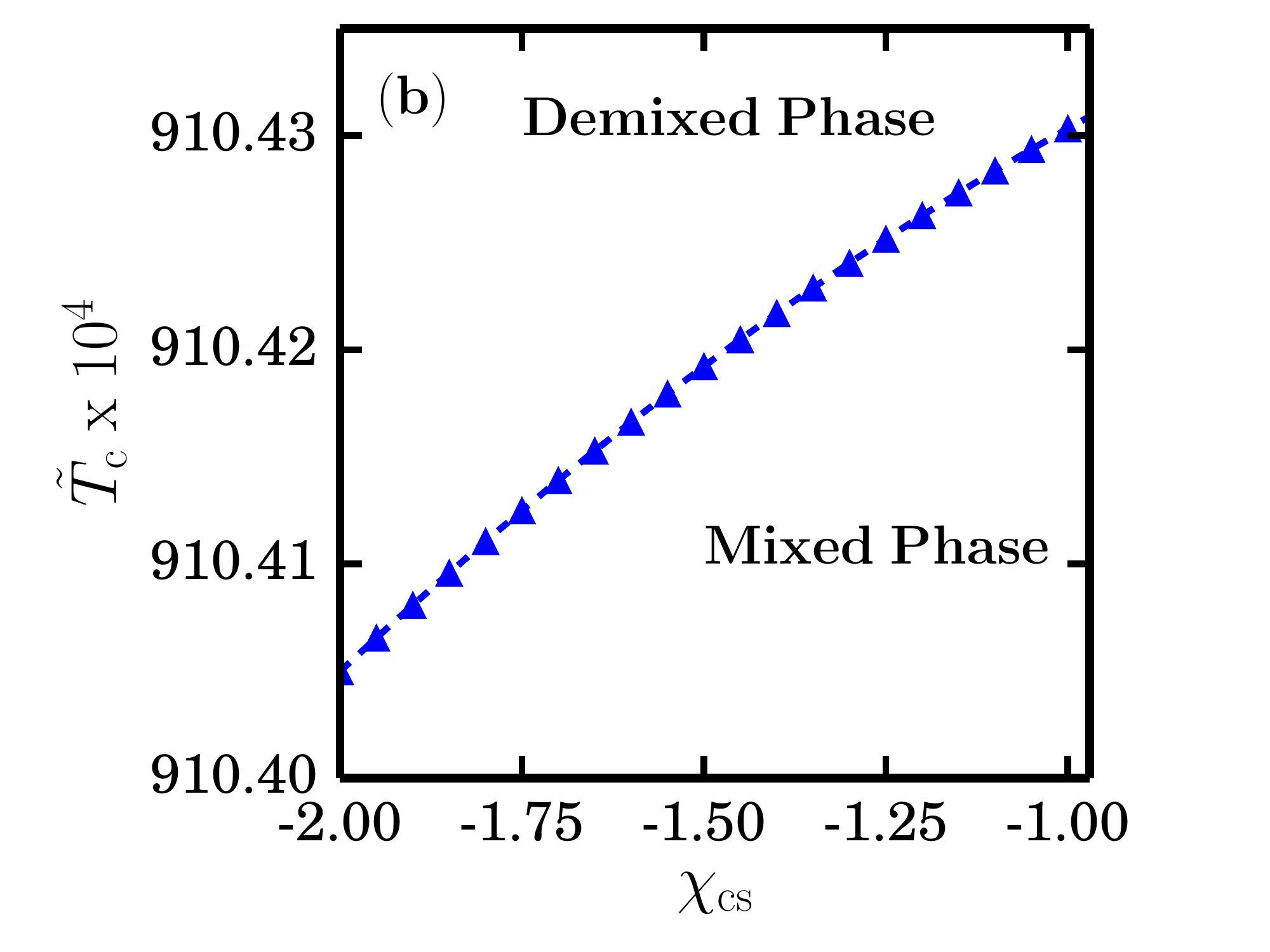}}
\end{center}
\caption{(a)Variation of the determinant of the hessian matrix with $\chi_{\rm ps}$ for different $\chi_{\rm cs}$ values. (b) Variation of the transition temperature $\tilde{T}_{\rm c}$ with $\chi_{\rm cs}$. Both results are for the multiple chain system (Eq.~(\ref{eq:multichain})). The calculations were performed at $N=10000$, $\chi_{\rm pc}=0$, $\phi_{\rm p}=0.01$ and $\phi_{\rm c}=0.01$}
\label{fig:bulk}
\end{figure}

\begin{figure}[h]
\begin{center}    
\includegraphics[scale=0.35]{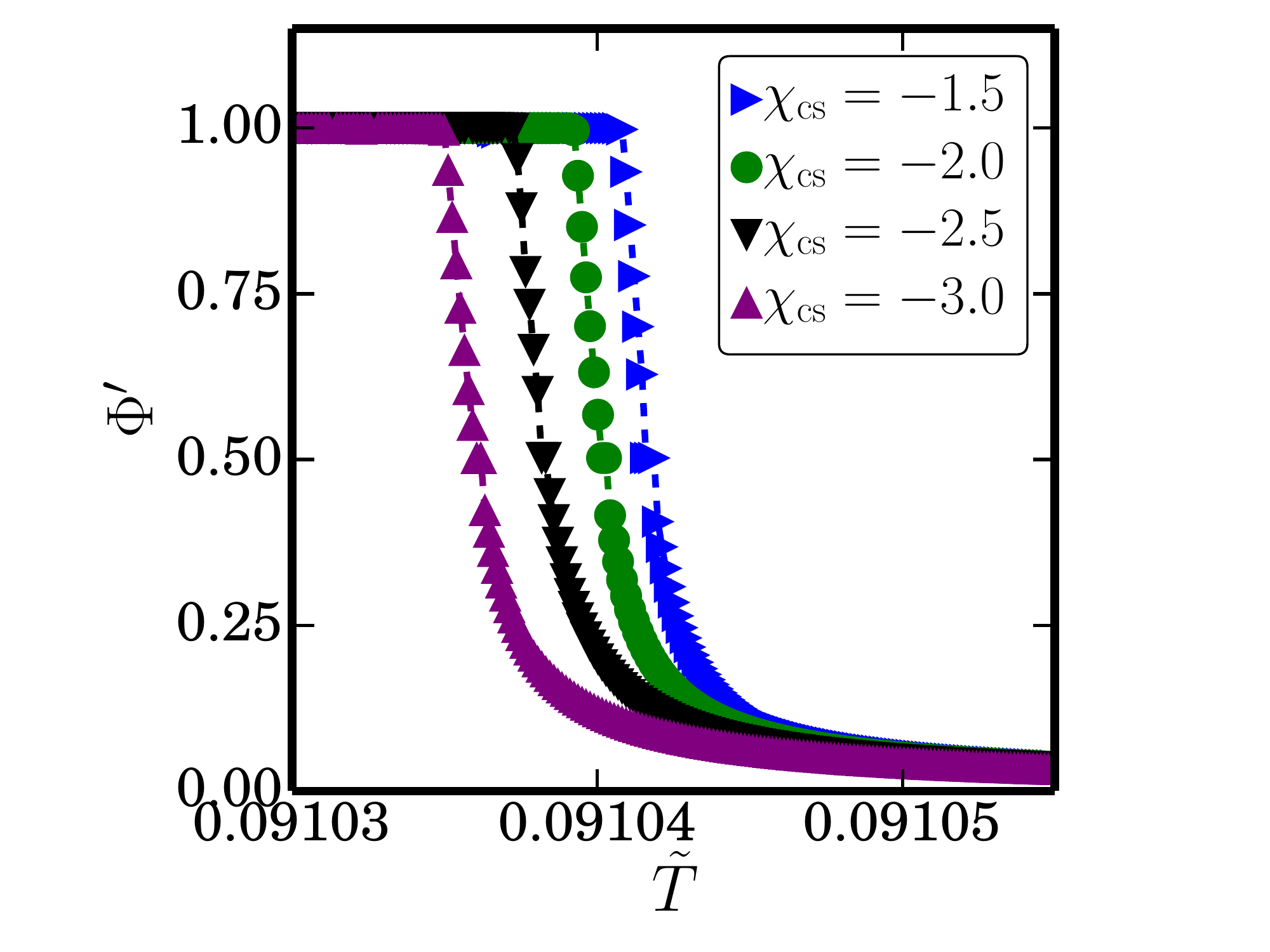}
\end{center}
\caption{Variation of $\Phi^{'}$ with temperature $\tilde{T}$ for different $\chi_{\rm cs}$ in the single chain system (Eq.~(\ref{eq:freenergy})). The calculations were performed at $N=10000$, $\chi_{\rm pc}=0$, $\Phi_{\rm p}=0.01$ and $\Phi_{\rm c}=0.01$.}
\label{fig:bb1}
\end{figure}

The mean-field phase behavior of the multiple chain system (Eq.~(\ref{eq:multichain})) which has been obtained using the procedure in Sec.~\ref{sec:mchain} is shown in Fig.~\ref{fig:bulk}. Figure~\ref{fig:ba} shows the variation of $D$ with $\chi_{\rm ps}$ for different $\chi_{\rm cs}$. The value of the polymer-solvent interaction parameter at the spinodal $\chi_{\rm ps}^{\rm spi}$, is lowered with decrease in $\chi_{\rm cs}$ which indicates a decrease in the solubility of the polymer. The monotonic decrease of the corresponding transition temperature $\tilde{T}_{\rm c}$ with  decrease in  $\chi_{\rm cs}$ can be seen in Fig.~\ref{fig:bb}.
This behavior can be understood by examining  the overall second virial coefficient, which is given by the following expression (see SI for details), 
 \begin{equation}\label{eq:virialcoeff}
 \begin{split}
 B&=1-\left[2(1-X_{\rm c})\chi_{\rm ps}(\tilde{T}) + 2X_{\rm c}\chi_{\rm pc} - 2X_{\rm c}(1-X_{\rm c})\chi_{\rm cs}\right]\\&\approx1-2\left[\chi_{\rm ps}(\tilde{T}) - X_{\rm c}\chi_{\rm cs}\right].
\end{split}
 \end{equation}
The value of $\chi_{\rm ps}$ at which $B=0$ is lowered as $\chi_{\rm cs}$ decreases. This indicates that temperature of phase separation $\tilde{T}_{\rm c}$ drops with a decrease in the bulk solvent mixture. The variation of the coil-to-globule transition with $\chi_{\rm cs}$  in the single chain system (see Sec.~\ref{sec:singlechain}) is shown in Fig.~\ref{fig:bb1}. The coiled state is indicated by $\Phi^{'}=1$ and the globule state by a lower value of $\Phi^{'}$ ($\Phi^{'}\neq 0$ as $\overline{R}_{\rm g}$ is finite). The transition temperature decreases with a decrease in $\chi_{\rm cs}$. Another point to note is that the transition temperature  is the same as in the multiple chain calculations (Fig.~\ref{fig:bb}) which indicates the coexistence of the bulk phase separation with the coil-to-globule transition. This feature may be expected due to the nature of the mean-field theory and has been experimentally observed  in PNiPAM water-alcohol systems at low alcohol concentrations.~\cite{Scherzinger2010, Bischofberger2014a}

To understand the results from a phenomenological point of view, let us first start with the  free energy difference of a solvent molecule between the BS and US states in the absence of cosolvent;
\begin{equation}\label{eq:deltaf}
\Delta F = F_{\rm BS}-F_{\rm US}=U_{\rm BS}-U_{\rm US} - T(S_{\rm BS}-S_{\rm US}),
\end{equation}
where $U_{\rm BS}$ ($U_{\rm US}$) and $S_{\rm BS}$ ($S_{\rm US}$) are the 
energy and the entropy of the BS (US) state of the solvent, respectively. Using the model parameters defined in Sec.~\ref{sec:models}, the above quantity can be expressed as 
\begin{equation}
\Delta F=-(u-w) +k_{\rm B}T \ln{(q-1})=k_{\rm B}T\ln x,
\label{eq:DeltaF}
\end{equation}
where
\begin{equation}\label{eq:eqx}
  x=(q-1)\exp{\left(-\frac{u-w}{k_{\rm B}T}\right)}.
\end{equation}
The phase transition occurs when $x>1$. In our previous study,\cite{Bharadwaj2017} we showed that the LCST in the case of pure solvent is dependent on the competition between the mean energy difference between the bulk and the bound solvent ($U_{\rm BS}-U_{\rm US}$), and the bound solvent entropy loss ($S_{\rm BS}-S_{\rm US}$). For $x<1$, the enthalpic gain $(u-w)$ dominates (mixed state), whereas for $x>1$, the bound solvent entropy loss $\ln{(q-1)}$ dominates (demixed state). The addition of cosolvent leads to a decrease in the energy of the bulk solvent $u$. The energy of the bound solvent $w$, on the other hand, remains unaffected as the cosolvent prefers to stay in bulk. The expression for $x$ in Eq.~(\ref{eq:eqx}) takes the following form in the case of solvent-cosolvent mixtures,
\begin{equation}\label{eq:xmix}
  x=(q-1)\exp{\left(-\frac{u_{\rm m}(X_{\rm c}, \chi_{\rm cs})-w}{k_{\rm B}T}\right)},
\end{equation}
where $u_{\rm m}$ is the energy of the solvent in the solvent-cosolvent mixture, which depends on $X_{\rm c}$ and $\chi_{\rm cs}$.  As mentioned earlier, the energy of the solvent in the bulk $u_{\rm m}$, decreases with the addition of the cosolvent. This reduces the mean energy difference between the bulk and the bound solvent leading to a decrease in the LCST. The addition of cosolvent diminishes the energetic advantage of the bound solvent by decreasing the energy of the bulk solvent.  Hence, it can be said that the decrease in LCST with the addition of cosolvent is driven by the decrease in the bulk solvent mean energy. This is in agreement with the simulation results and is supported by the experimental work of Bischofberger et al.\cite{Bischofberger2014,Bischofberger2014a} To highlight the influence of cosolvent, $x$ in Eq.~(\ref{eq:xmix}) can be rewritten in the following manner, 
\begin{widetext}
\begin{equation}
  x=(q-1)\exp{\left(-\frac{u_{\rm m}(X_{\rm c}, \chi_{\rm cs})-w}{u-w}\ \ \frac{u-w}{k_{\rm B}T}\right)}=(q-1)\exp{\left(-\frac{1}{\gamma(X_{\rm c}, \chi_{\rm cs})\tilde{T}}\right)},
\end{equation}
\end{widetext}
where 
\begin{equation}\label{eq:gamma}
\gamma(X_{\rm c}, \chi_{\rm cs})=\frac{u-w}{u_{\rm m}(X_{\rm c}, \chi_{\rm cs})-w},
\end{equation}
 is the ratio of the mean energy difference between the bulk and the bound solvent in the case of pure solvent to that in the solvent-cosolvent mixture. The quantity $\gamma$ characterizes the extent of decrease of LCST with cosolvent addition and $\gamma=1$ corresponds to the case of pure solvent. High values of $\gamma$ indicate a higher decrease in bulk solvent energy, which in turn indicates a larger drop in the LCST in comparison to the pure solvent case. 

Another aspect which has to be examined is the hydrophilicity of the polymer. Let us consider the case of a polymer where the bound solvent energy is $w'$ and look at the expression in Eq.~(\ref{eq:xmix}),
\begin{widetext}
\begin{equation}\label{eq:xmixcosol}
\begin{split}
  x&=(q-1)\exp{\left(-\frac{u_{\rm m}(X_{\rm c}, \chi_{\rm cs})-w'}{k_{\rm B}T}\right)}\\&=(q-1)\exp{\left(-\frac{u_{\rm m}(X_{\rm c}, \chi_{\rm cs})-w'}{u_{\rm m}(X_{\rm c}, \chi_{\rm cs})-w}\ \ \frac{u_{\rm m}(X_{\rm c}, \chi_{\rm cs})-w}{u-w}\ \ \frac{u-w}{k_{\rm B}T}\right)}\\&=(q-1)\exp{\left(-\frac{\lambda}{\gamma\tilde{T}}\right)}=(q-1)\exp{\left(-\frac{\Delta_{u_{\rm m},w'}}{\tilde{T}}\right)}
  \end{split}
\end{equation}
\end{widetext}
where 
\begin{equation}
\lambda=\frac{u_{\rm m}(X_{\rm c}, \chi_{\rm cs})-w'}{u_{\rm m}(X_{\rm c}, \chi_{\rm cs})-w},
\end{equation} 
 is the ratio of mean energy difference between the bulk and bound solvent for a polymer with the respect to the base polymer (bound solvent energy $w$) and 
\begin{equation}\label{eq:delta}
\Delta_{u_{\rm m},w'}=\frac{\lambda}{\gamma},
\end{equation}
 is the parameter which captures the cumulative effect of the cosolvent and the polymer hydrophilicity on $\tilde{T}_{\rm c}$, and $\gamma$ is defined in Eq.~(\ref{eq:gamma}). Higher values of $\lambda$ ($>1$) indicate higher hydrophilicity of the polymer. The LCST of the system will drop with the decrease in the value of $\Delta_{u_{\rm m},w'}$. For a fixed solvent-cosolvent mixture (fixed $X_{\rm c}, \gamma$), as the hydrophilicity of the polymer rises, the mean energy difference between the bound solvent and the bulk solvent increases due to increase of $\lambda$ and $\Delta_{\rm u_{\rm m},w'}$. This enhances the stability of the bound solvent which reduces the extent of LCST decrease induced by the cosolvent. Hence, to obtain the same drop in LCST, one requires a larger increase in $\gamma$  as the hydrophilicity of the polymer increases. This is in agreement with the results in the simulation studies.
\subsection{Comparison with experiments}
\begin{figure}[h]
\begin{center}    
\includegraphics[scale=0.35]{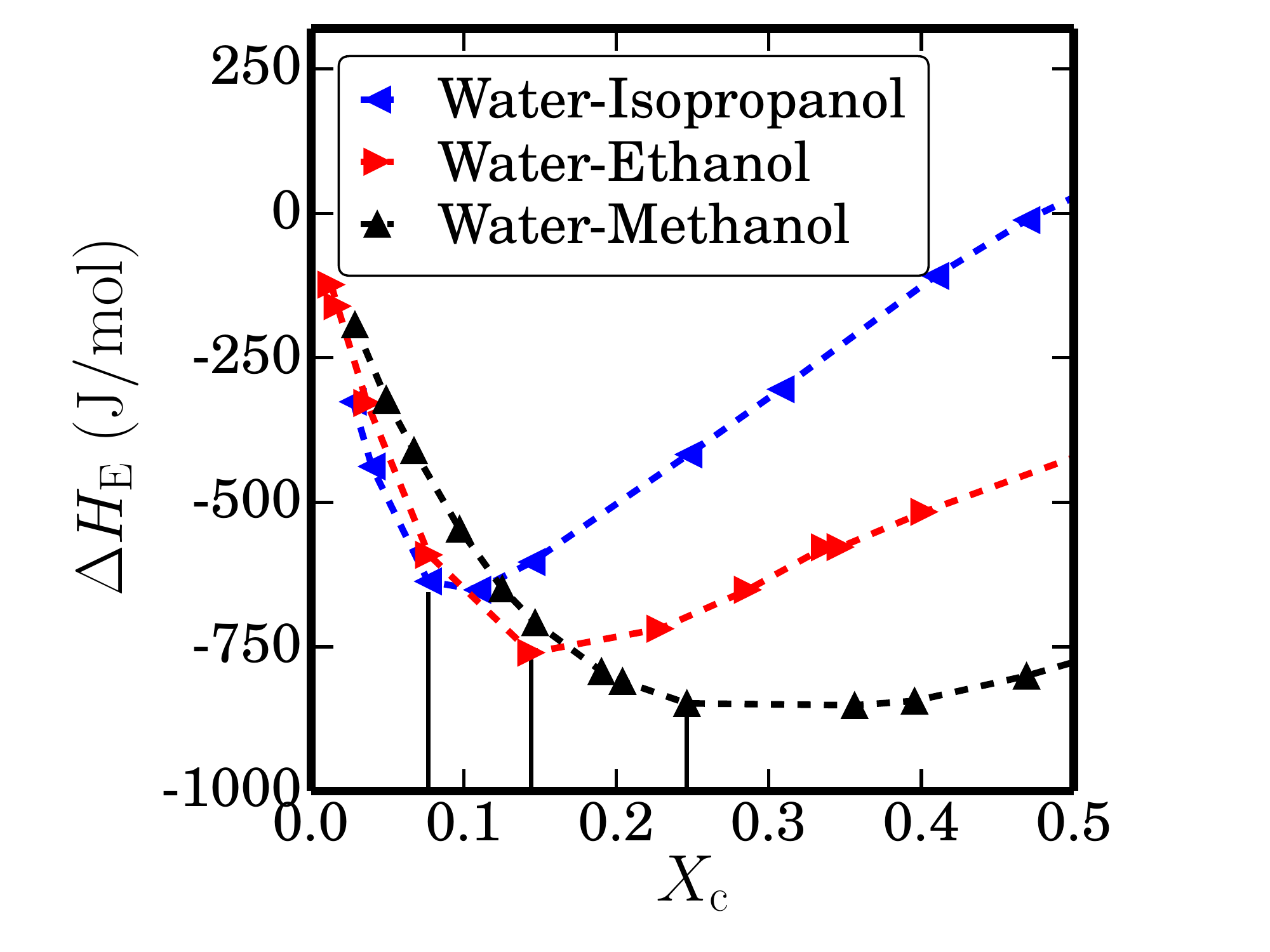}
\end{center}
\caption{Enthalpy of mixing in water-alcohol (kosmotropic or order-making) mixtures as a function of alcohol concentration $X_{\rm c}$ (=$x_{\rm alcohol}/(x_{\rm alcohol}+x_{\rm water}$)), at 25$\degree$C. Data taken from Lama and Lu.\cite{Lama1965} The solid vertical lines are the concentrations ($X_{\rm c}^{*}$) at which $\Delta H_{\rm E}$ is minimum for the respective alcohols. The low concentration regime is $X_{\rm c}<X_{\rm c}^{*}$.}
\label{fig:methe}
\end{figure}

\begin{figure}[h]
\begin{center}    
\subfigure{\label{fig:ca}\includegraphics[scale=0.35]{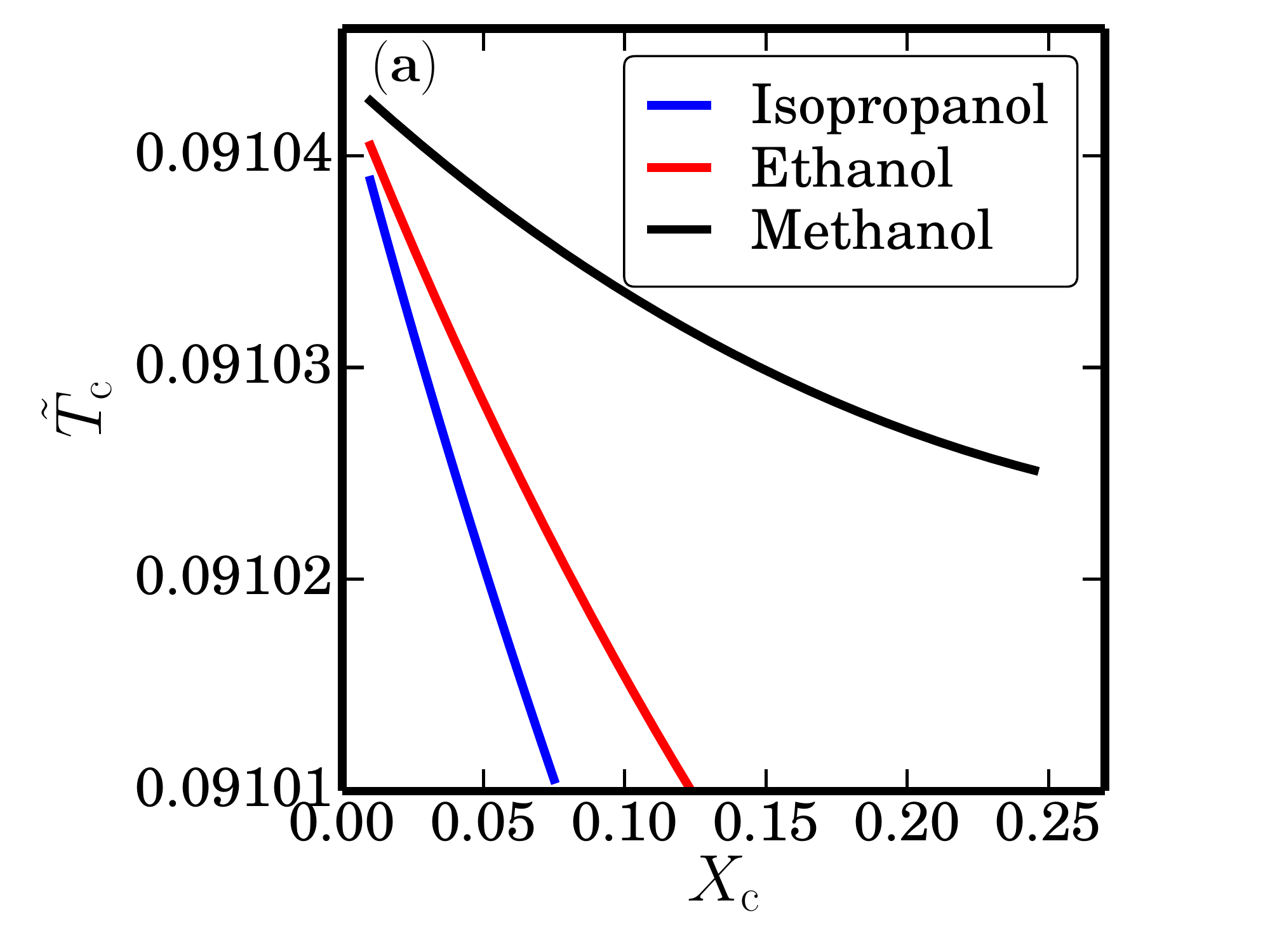}}
\subfigure{\label{fig:cb}\includegraphics[scale=0.35]{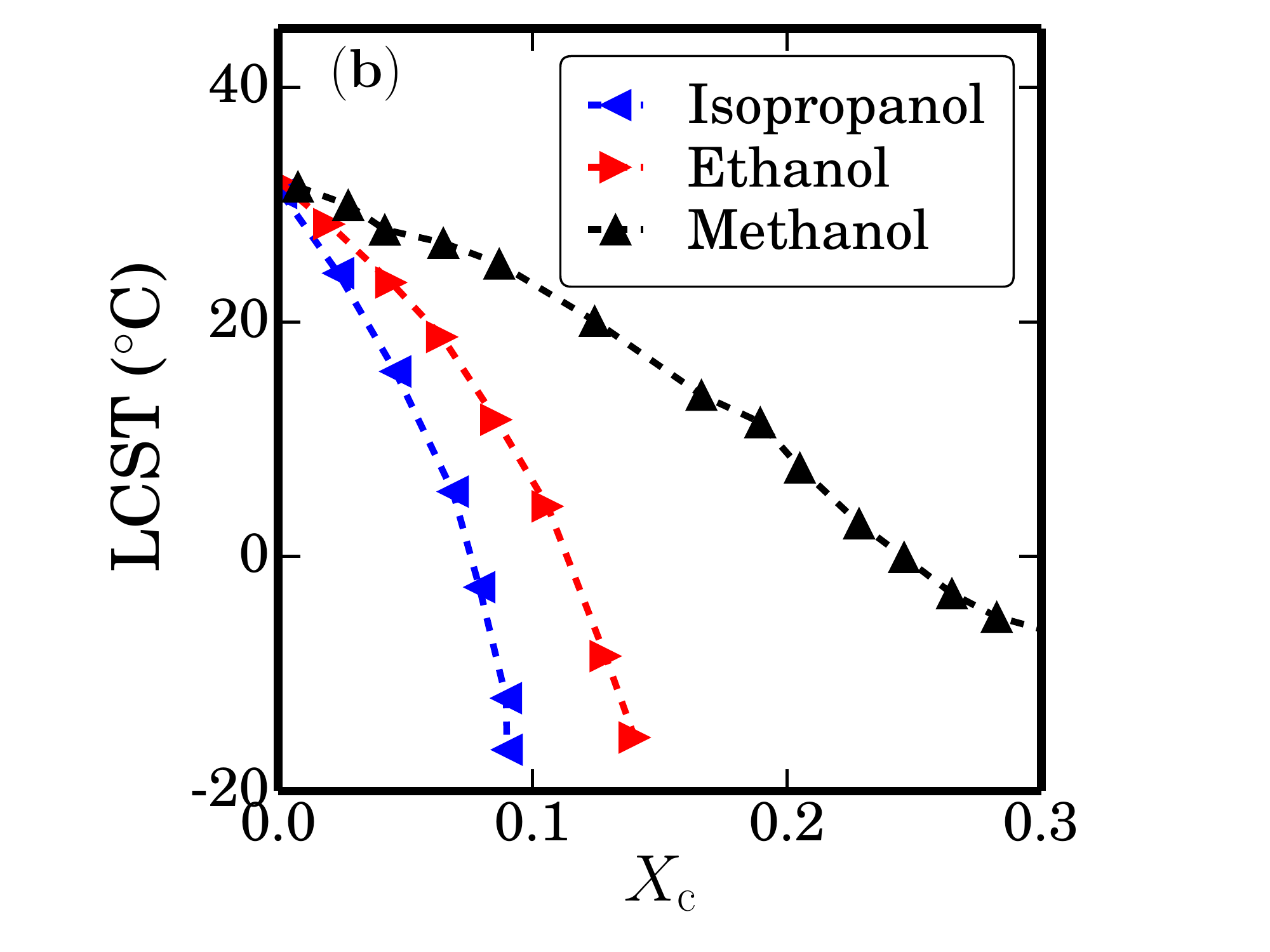}}
\end{center}
\caption{Variation of transition temperature ($\tilde{T}_{\rm c}$ or LCST) with $X_{\rm c}$ for different alcohols. (a)Theoretical prediction  in the multiple chain framework, (b) Experimental data for PNiPAM.\cite{Lama1965}} 
\label{fig:conc}
\end{figure}

Let us first look at the LCST decrease with the addition of different alcohols. Figure~\ref{fig:methe} shows the variation of  the excess enthalpy of mixing, $\Delta H_{\rm E}$ with $X_{\rm c}$ for different water-alcohol mixtures. Here $X_{\rm c}^{*}$ is the concentration at which $\Delta H_{\rm E}$ is minimum and the low concentration regime is defined by $X_{\rm c}<X_{\rm c }^{*}$. It is important to note that the decreasing (and negative) trends of $\Delta H_{\rm E}$ and relatively constant values of $X^{*}_{\rm c}$  are  observed over the temperature range under consideration.\cite{Lama1965,Larkin1975,Peeters1993} This data ($X_{\rm c }<X_{\rm c }^{*}$) is fitted to Eq.~(\ref{eq:chics})  to obtain $\chi_{\rm cs}$ for different alcohols (see Table.~\ref{table:table1}).
These $\chi_{\rm cs}$ values were incorporated in the multiple chain framework (Eq.~(\ref{eq:multichain})) to obtain the theoretical prediction of LCST with $X_{\rm c}$ for different alcohols. 
\begin{table}[h]
\footnotesize
 \caption{$\chi_{\rm cs}$ and $X_{\rm c}^{*}$ for different alcohols in the low $X_{\rm c}$ limit.}
 \begin{tabular}{|c|c|c|c|}
 \hline
    & Isopropanol& Ethanol& Methanol\\
    \hline
   $\chi_{\rm cs}$&-2.47&-2.00&-1.18\\
   \hline
   $X_{\rm c}^{*}$&0.076&0.144&0.246\\
   \hline
 \end{tabular}
 \label{table:table1}
\end{table}
Figure~\ref{fig:ca} shows the variation of $\tilde{T}_{\rm c}$ (LCST) with $X_{\rm c}$ for alcohols. It can be observed that $\tilde{T}_{\rm c}$ decreases with an increase in alcohol concentration. Additionally, the extent of $\tilde{T}_{\rm c}$ decrease (per unit concentration) is highest for isopropanol followed by ethanol and methanol. These results qualitatively match with the experimental data shown in Fig.~\ref{fig:cb} for the PNiPAM, water and alcohols systems. The agreement between experiment data and theoretical prediction indicates that the mean energetics of the solvent-cosolvent mixtures are more important than the structural details for understanding the phenomena.
In the simulation and theoretical studies, we have shown that the LCST decrease is dependent on the interplay between the polymer hydrophilicity and the extent of cosolvent induced decrease in bulk solvent energy. This can be used to explain the absence of the LCST decrease in PDEA, water and methanol mixtures. To understand this in a better way, let us first look at the case of PDEA and PNiPAM in pure water. The non-polar solvent accessible surface area (SASA) for PDEA is higher than PNiPAM.\cite{Nayar2017,Dalgicdir2017} This implies that the bound solvent energy (bound solvent entropy loss) is lower (higher) for PDEA in comparison to PNiPAM ($(\overline{\epsilon}_{\rm \textsc{as}})_{\rm PDEA}> (\overline{\epsilon}_{\rm \textsc{as}})_{\rm PNiPAM}$ or $(w)_{\rm PDEA}<(w)_{\rm PNiPAM},\ q_{\rm PDEA} > q_{\rm PNiPAM}$).\cite{Muller1990,Grdadolnik2017} In other words, the mean energy difference between the bulk and bound solvent ($\Delta \overline{\epsilon}=\overline{\epsilon}_{\rm \textsc{as}}-\overline{\epsilon}_{\rm ss}$ or $\lambda$), and the entropy loss of the bound solvent are higher for PDEA than PNiPAM ($\Delta \overline{\epsilon}_{\rm PDEA}>\Delta \overline{\epsilon}_{\rm PNiPAM}$ or $\lambda_{\rm PDEA}>\lambda_{\rm PNiPAM}$). This might be the reason for the almost same LCST in aqueous solutions of PDEA and PNiPAM. 
Given the higher $\Delta \overline{\epsilon}$ in PDEA in comparison to PNiPAM, the drop in the energy of the bulk solvent ($\overline{\epsilon}_{\rm \textsc{ss}}-\overline{\epsilon}_{\rm bulk}$ or $\gamma$)  with addition of methanol is not sufficient to induce a LCST decrease. 

\begin{figure}[h]
\begin{center}    
\subfigure{\label{fig:ca3}\includegraphics[scale=0.35]{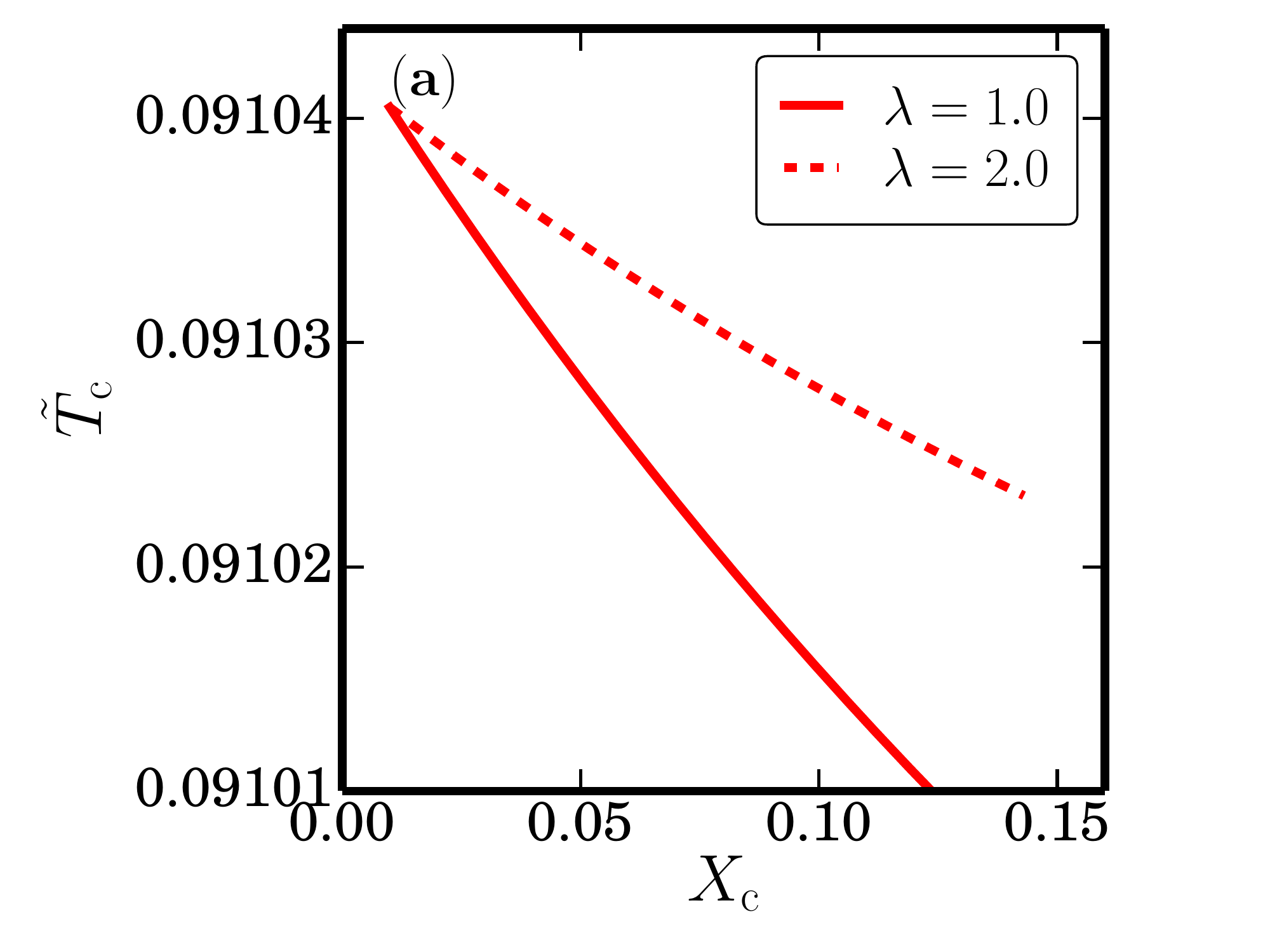}}
\subfigure{\label{fig:cb3}\includegraphics[scale=0.35]{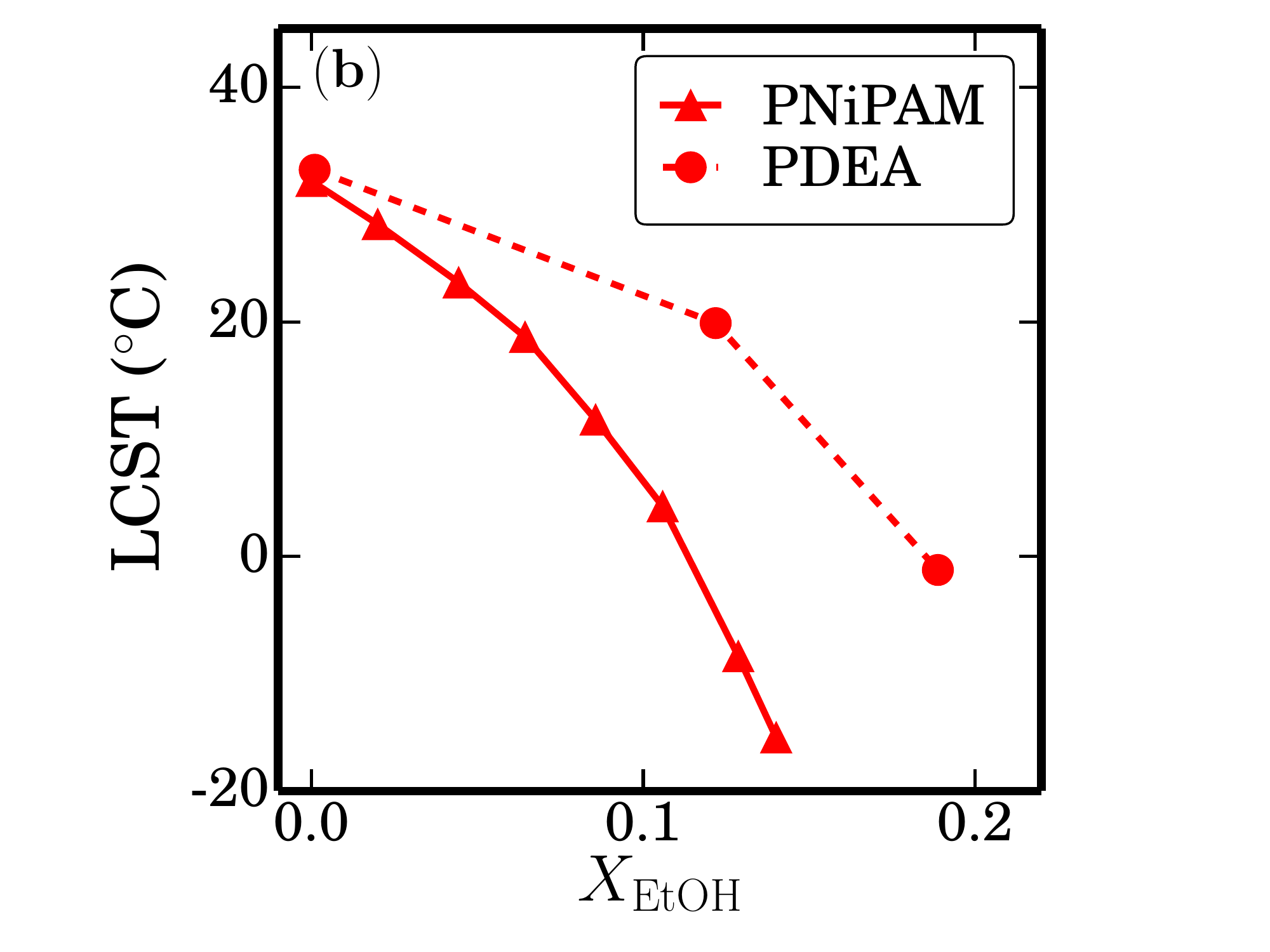}}
\end{center}
\caption{(a)Variation of transition temperature $\tilde{T}$ with $X_{\rm c}$ for varying polymer hydrophilicity at $\chi_{\rm cs}=-2.00$ (Ethanol) in the multiple chain framework. (b) Experimental variation of LCST with ethanol concentration for PNiPAM\cite{Bischofberger2014} and PDEA\cite{Maeda2002}.}
\label{fig:rt4}
\end{figure}

From the simulation point of view, PNiPAM and PDEA can be considered analogous to polymers with $\overline{\epsilon}_{\rm \textsc{as}}=1.7$ and $\overline{\epsilon}_{\rm \textsc{as}}=2.0$, respectively. Then, it can be seen from Fig.~\ref{fig:rt3_3} that a cosolvent with $\overline{\epsilon}_{\rm bulk}=1.05$ (analogous to methanol) is able to induce a change of $\overline{R}_{\rm g}$ in  $\overline{\epsilon}_{\rm \textsc{as}}=1.7$ (PNiPAM) but not in $\overline{\epsilon}_{\rm \textsc{as}}=2.0$ (PDEA). In the multiple chain theory, this can be understood by considering the parameter $\Delta_{u_{\rm m}, w'}$ (Eq.~(\ref{eq:delta})). In the case of PNiPAM, the addition of methanol leads to a decrease in $\Delta_{u_{\rm m}, w'}$ (increase in $\gamma$) which causes a drop in LCST. In PDEA, however, this $\gamma$ increase is countered by the increase in $\lambda$ (higher hydrophilicity) due to which $\Delta_{u_{\rm m}, w'}$ does not change which in turn leads to a constant LCST.
This suggest that cosolvents such as ethanol, isopropanol, which induce a much stronger decrease in the bulk solvent energy are required to observe an LCST decrease in the case of PDEA. An additional point which arises from the above discussion is that for a given cosolvent, the extent of decrease in LCST is lower for polymers with higher hydrophilicity. Figure~\ref{fig:ca3} shows the variation of the LCST with $X_{\rm c}$ for different polymer hydrophilicity ($\lambda$) at $\chi_{\rm cs}=-2.00$ (ethanol). The model assumes that the polymers have the same LCST in pure solvent ($X_{\rm c}=0$). The extent of LCST change decreases at higher $\lambda$. This is qualitatively similar to the experimental variation of the LCST for PNiPAM (analogous to $\lambda=1.0$) and PDEA ($\lambda=2.0$) with increase in ethanol concentration (Figure~\ref{fig:cb3}).  This is also supported by the simulation results in Fig.~\ref{fig:rt3}.

\begin{figure}[h]
\begin{center}    
\includegraphics[scale=0.35]{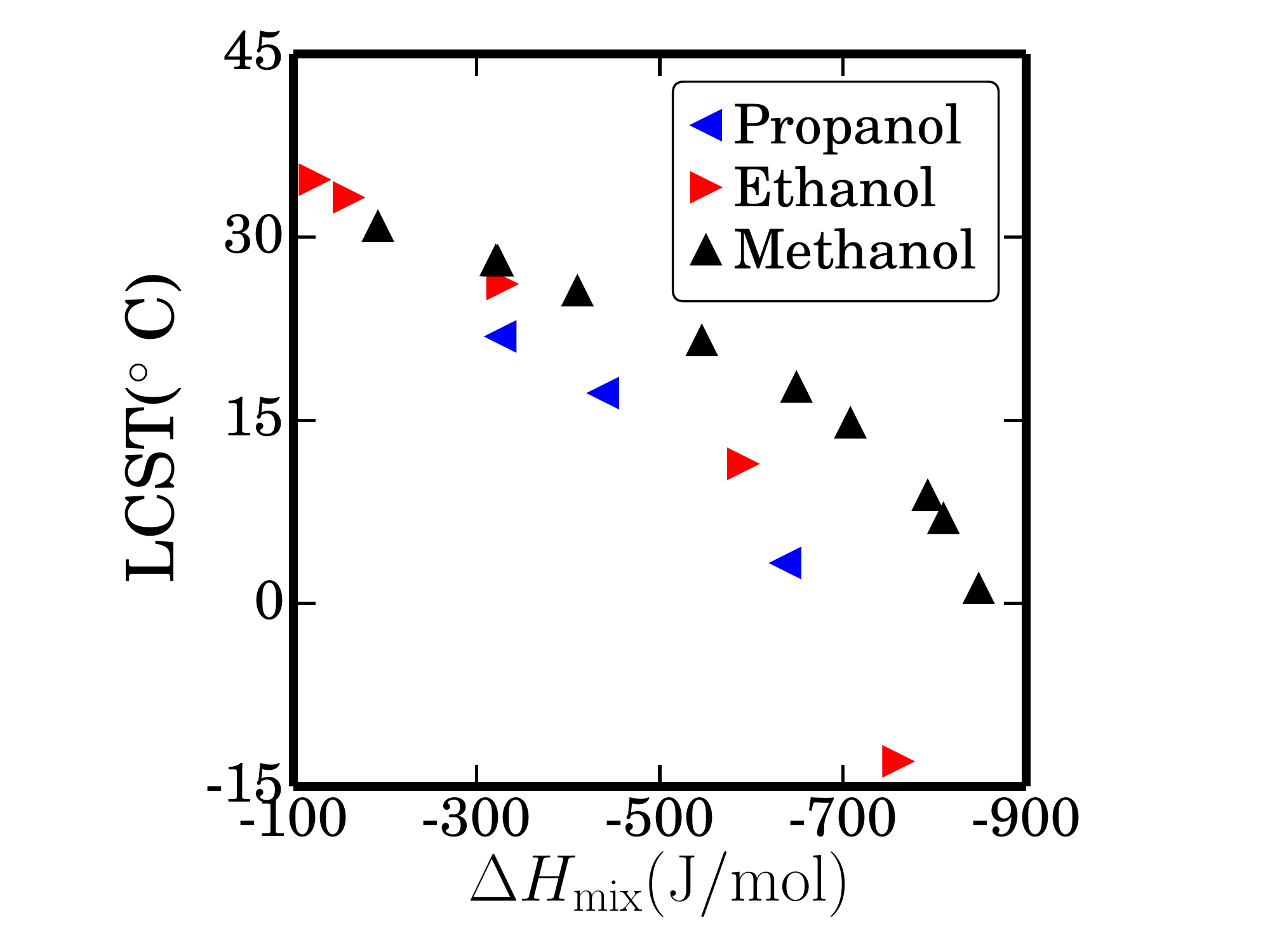}
\end{center}
\caption{Experimental variation of LCST with the excess enthalpy of mixing for different alcohol-water mixtures.\cite{Lama1965,Bischofberger2014}}
\label{fig:rt6}
\end{figure}

A point to note is that the temperature and its extent of decrease depend on the model parameters  namely the number of solvent orientations in the KW model, $q$, and the degree of polymerization, $N$. By suitable tuning of these parameters, the extent of change in temperature  can be varied to a limited extent. The temperature variation can be increased by reducing the value of $q$ to small values. We desist from going to smaller values of $q$  as it is  not realistic within the  framework of the one dimensional Kolomeisky-Widom model, which assumes $q$ to be large in the analysis. Due to this, the inferences from our theoretical model are based only on the trends and not the quantitative estimates. However,  Barkema and Widom\cite{Barkema2000} have shown through computer simulations that the functional form of the potential remains the same for the two and the three dimensional cases and it is applicable for small $q$ values as well. This indicates that the trends from our models are consistent at all values of $q$. Another point of observation is that there is a difference in the curvature of the theoretical and experimental trends. This may be due to the fact that $\chi_{\rm cs}$ is assumed to be independent of the cosolvent concentration in the model. In real liquid mixtures, $\chi_{\rm cs}$ is dependent on the cosolvent composition. This variation of $\chi_{\rm cs}$ will differ for each solvent-cosolvent combination. In case of kosmotropic cosolvents, the strength of the water hydrogen bonded network increases with cosolvent concentration ($X_{\rm c}<X_{\rm c}^{*}$) which indicates that the magnitude of $\chi_{\rm cs}$ may be an increasing function of $X_{\rm c}$. However, this variation is dependent on the chemical specific details, which is not within the scope of this paper.  However, given the qualitative matching of the trends in the models with the experimental data for the acrylamide family of polymers, it can be said that these system specific details may not affect the overall generic mechanism. This can be seen by looking at the variation of the experimentally observed LCST in PNiPAM, water and alcohols mixtures with the excess enthalpy of mixing of the water-alcohol mixture. A weak dependence of the LCST on the cosolvent chemical details  in Fig.~\ref{fig:rt6}  supports our hypothesis that the decrease of the mean energy of the bulk solvent-cosolvent mixture is the dominant contribution. 

\section{Summary and Conclusions}
In this paper, we have  examined the effects of cosolvents on the LCST in thermoresponsive polymer solutions. Our approach has been to address the phenomenon using a generic bead-spring simulations and theory such that dominant interactions leading to the phenomena can be identified. In the introduction of this paper, we mentioned that the from a generic point of view, the phenomenon of cononsolvency can be classified on the basis of cosolvent type and concentration regime.  The question which arises is that whether there can be a generic explanation for the cononsolvency phenomenon which is independent of concentration regime and cosolvent type. Our observations are contrary to such an explanation. As the concentration of the cosolvent exceeds $X_{\rm c}^{*}$, the change in LCST becomes dependent on the molecular weight, concentration and architecture. Additionally, there is no temperature dependent coil-to-globule transition accompanying the bulk phase separation. Also, the exclusion of cosolvent from the polymer hydration shell is not valid. In this regime the cosolvent-polymer interaction may be dependent on the specific chemical details as in the case of PNiPAM and PDEA in urea-water mixtures.

In the current work, we have focussed on kosmotropic (order-making) cosolvents in the low concentration limit. In the simulation studies, we have considered a single bead-spring polymer chain in an explicit solvent-cosolvent mixture. The interaction parameters are independent of temperature. In our theoretical models, we have considered both multiple polymer chain system and single polymer chain in a solvent-cosolvent mixture. The interactions between the different components have been modeled by Flory-Huggins interaction parameters. The polymer-solvent interaction parameter, $\chi_{\rm ps}$ is modeled through the solvophobic potential, which follows from our earlier work\cite{Bharadwaj2017} on the LCST of thermoresponsive polymers in a pure solvent.\\
\indent Our calculations indicate that the LCST decrease with addition of kosmotropic cosolvents may be driven by the decrease in the energy of the bulk solvent. The LCST  is dependent  on the mean energy difference between the bound and the bulk solvent. This reduces the stability of the bound solvent leading to the decrease in the LCST. These results are in qualitative agreement with the experimental work on PNiPAM of Bischofberger et al.~\cite{Bischofberger2014a} Additionally, we have shown that the cosolvents which have a weak decrease in mean bulk solvent energy may not be able to induce the LCST decrease as the hydrophilicity of the polymer increases.  This is in agreement with the experimental observation that the LCST of aqueous PDEA is not modified with the addition of methanol. PDEA is more hydrophilic (higher mean energy difference between bound and bulk) than PNiPAM and the lowering of the bulk solvent energy by methanol is not sufficient to affect the LCST in this case. This shows that our simulation and theoretical models are able to explain the trends qualitatively across the acrylamide family of thermoresponsive polymers. An important point to emphasize here is that the simulation model is able to exhibit both the temperature dependent and cosolvent dependent transitions without the presence of temperature dependent interaction parameters. The qualitative matching of our  predictions with experimental data indicates that the mean energetics of the system are the dominant considerations in the system. Therefore, from a phenomenonlogical point of view, it can be said that details regarding the structure of the cosolvent and its interaction with the solvent may not be very important in comparison to the mean energetics. From an application point of view, this indicates that the choice of kosmotropic cosolvent to control the LCST can be made by examining its effect on the bulk energy. 
\label{sec:sum}
\section{Acknowledgments}
The computations were carried out at the High Performance Computing Facility at IIT Madras. S.K.\ acknowledges support from a Grant-in-Aid for Scientific Research on
Innovative Areas ``\textit{Fluctuation and Structure}" (Grant No.\ 25103010)
from the Ministry of Education, Culture, Sports, Science, and Technology of
Japan and from a Grant-in-Aid for Scientific Research (C) (Grant No.\
15K05250) from the Japan Society for the Promotion of Science (JSPS). S.B.\  acknowledges Tokyo Metropolitan University for the support provided
through the co-tutorial program. 
\bibliography{references}
\newpage
\section{Supplementary Material}
\section{Second virial coefficient $B_{\rm KW}$ : Kolomeisky-Widom potential}
\label{sec:kolomeiskywidom}
The second virial coefficient of the Kolomeisky-Widom model can be calculated using the following expression,
\begin{equation}
B_{\rm KW}=4\pi\int_{0}^{\infty} {\rm d}r \, r^2 \left(1-e^{-E(r)/k_{\rm B}T}\right), 
\label{eq:second_virial}
\end{equation}
where the prefactor of $1/2$ is included within $B_{\rm KW}$ and $E(r)$ is the monomer-monomer interaction potential which has the  following form:
\begin{equation}\label{eq:interaction}
E(r) = \left\{
\begin{array}{ll}
\infty &: r<\sigma \\
U(r) &: r>\sigma 
\end{array}
\right..
\end{equation} 
Here $E(r)$ for $r<\sigma$ corresponds to the (hard-core) excluded volume 
interaction, and $U(r)$ for $r>\sigma$ is the KW solvophobic potential. The expression for $U(r)$ is
\begin{equation}\label{eq:widom_potential}
U(r) = -k_{\rm B}T\ln{\left[1+\left(\frac{1+Q}{1-Q}\right)
\left(\frac{1-S}{1+S}\right)^{(r-\sigma)/\sigma}
\right]},
\end{equation} 
where
\begin{equation}\label{eq:6}
S=\left[1-\frac{4 x}{(1+x)^2}\left(1-\frac{1}{c}\right)\right]^{1/2},
\end{equation}
\begin{equation}\label{eq:7}
Q=\frac{\rm \sgn{({\it x}-1)}}{\left[1+4 x/(x-1)^2 c\right]^{1/2}}, 
\end{equation}
\begin{equation}\label{eq:8}
x=\frac{q-1}{c}
\end{equation}
\begin{equation}\label{eq:9}
c=e^{(u-w)/k_{\rm B}T}.
\end{equation}
By using the above expression, $B_{\rm KW}$ is obtained  as follows
\begin{widetext}
\begin{equation}\label{eq:appvirial}
B_{\rm KW}=\frac{4\pi \sigma^{3}}{3}\left[1+ 3\left(\frac{1+Q}{1-Q}\right)\left( \frac{1}{\ln{L}} - 
\frac{2}{\left(\ln{L}\right)^{2}} +\frac{2}{\left(\ln{L}\right)^{3}} \right)\right],
\end{equation}
\end{widetext}
where $L=(1-S)/(1+S)$.  The dimensionless virial coefficient is given by
\begin{widetext}
\begin{equation}\label{eq:appvirial}
\tilde{B}_{\rm KW}(\tilde{T})=\left[1+ 3\left(\frac{1+Q}{1-Q}\right)\left( \frac{1}{\ln{L}} - 
\frac{2}{\left(\ln{L}\right)^{2}} +\frac{2}{\left(\ln{L}\right)^{3}} \right)\right],
\end{equation}
\end{widetext}
where
\begin{equation}
\tilde{T}=\frac{k_{\rm B}T}{(u-w)}.
\end{equation}
\section{Bulk phase separation}
\label{sec:bulkphase}
\subsubsection{Spinodal analysis}
The free energy per unit site for the multiple chain system is given by the following expression,
\begin{equation}\label{eq:mchain}
\frac{F}{k_{\rm B}T}=f=\frac{\phi_{\rm p}}{N}\ln{\phi_{\rm p}}+\phi_{\rm c}\ln{\phi_{\rm c}}+\phi_{\rm s}\ln{\phi_{\rm s}}+\chi_{\rm cs}\phi_{\rm c}\phi_{\rm s}+\chi_{\rm ps}\phi_{\rm p}\phi_{\rm s}+\chi_{\rm pc}\phi_{\rm p}\phi_{\rm c},
\end{equation}
the corresponding Hessian matrix is,
{\begin{equation}\label{eq:apphessian}
H=\left( \begin{array}{cc}
\frac{\partial^{2} f}{\partial \phi_{\rm p}^{2}} & \frac{\partial^{2}f}{\partial \phi_{\rm p}\partial \phi_{\rm c}}\\
\frac{\partial^{2} f}{\partial \phi_{\rm c}\partial \phi_{\rm p}} & \frac{\partial^{2} f}{\partial \phi_{\rm c}^{2}}  \end{array} \right),
\end{equation}}
where the elements of the matrix are as follows,
\begin{equation}
\begin{split}
\frac{\partial^{2} f}{\partial \phi_{\rm p}^{2}}=\frac{1}{N\phi_{\rm p}}+\frac{1}{1-\phi_{\rm c}-\phi_{\rm p}}-2\chi_{\rm ps},
\end{split}
\end{equation}
\begin{equation}
\frac{\partial^{2} f }{\partial \phi_{\rm p}\partial \phi_{\rm c}}=\frac{\partial^{2} f }{\partial \phi_{\rm c}\partial \phi_{\rm p}}=\frac{1}{1-\phi_{\rm c}-\phi_{\rm p}}-(\chi_{\rm cs}+\chi_{\rm ps}-\chi_{\rm pc})
\end{equation}
\begin{equation}
\begin{split}
\frac{\partial^{2} f}{\partial \phi_{\rm c}^{2}}=\frac{1}{\phi_{\rm c}}+\frac{1}{1-\phi_{\rm c}-\phi_{\rm p}}-2\chi_{\rm cs},
\end{split}
\end{equation}
The $\chi_{\rm ps}^{\rm spi}$ at the transition point ($D=0$) for a given $\phi_{\rm p}$ and $\phi_{\rm c}$ can be analytically obtained using the following expression,
\begin{widetext}
\begin{equation}\label{eq:appchips}
\begin{split}
\chi_{\rm ps}^{\rm spi}&=\frac{-1}{\phi_{\rm c}\phi_{\rm p}\phi_{\rm s}}\Bigg\{-\phi_{\rm p}\phi_{\rm s}\left[-1+\phi_{\rm c}(\chi_{\rm cs}+\chi_{\rm pc})\right]\\&\pm \Bigg(-\phi_{\rm p}\phi_{\rm s}[1-\phi_{\rm p}-2\phi_{\rm c}\phi_{\rm s}\chi_{\rm cs}][-n \phi_{\rm p}(1-2\phi_{\rm c}\chi_{\rm pc}) - \phi_{\rm c}]\Bigg)^{1/2}\Bigg\},
\end{split}
\end{equation}
\end{widetext}
where  $\phi_{\rm s}=1-\phi_{\rm p}-\phi_{\rm c}$.
\newpage
\subsubsection{Relation of $\chi_{\rm ps}$ to $\tilde{T}$}
The polymer-solvent Flory-Huggins parameter $\chi_{\rm ps}$ can be related to the virial coefficient $\tilde{B}_{\rm KW}$ (Eq.~(\ref{eq:appvirial})) by
\begin{equation}\label{eq:chipstot}
\begin{split}
\chi_{\rm ps}&=\frac{1}{2}-\tilde{B}_{\rm KW}\\
&=\left[1+ 3\left(\frac{1+Q}{1-Q}\right)\left( \frac{1}{\ln{L}} - 
\frac{2}{\left(\ln{L}\right)^{2}} +\frac{2}{\left(\ln{L}\right)^{3}} \right)\right],
\end{split}
\end{equation}
where $L=(1-S)/(1+S)$, 
\begin{equation}\label{eq:6}
S=\left[1-\frac{4 x}{(1+x)^2}\left(1-\frac{1}{c}\right)\right]^{1/2},
\end{equation}
\begin{equation}\label{eq:7}
Q=\frac{\rm \sgn{({\it x}-1)}}{\left[1+4 x/(x-1)^2 c\right]^{1/2}}, 
\end{equation}
\begin{equation}\label{eq:8}
x=\frac{q-1}{c}
\end{equation}
\begin{equation}\label{eq:9}
c=e^{1/\tilde{T}}.
\end{equation}
For a given value of $\chi_{\rm ps}$, the corresponding temperature $\tilde{T}$ (for fixed $q$) can be obtained by numerically solving Eq.~(\ref{eq:chipstot}). 

\newpage
\subsubsection{Overall second virial coefficient $B$}
To derive the expression for $B$ in the multiple chain system, let us first consider the volume fractions of the solvent and cosolvent in terms of $X_{\rm c}$ and $\phi_{\rm p}$
\begin{equation}
  \begin{split}
    \phi_{\rm c}&=X_{\rm c}(1-\phi_{\rm p}),\\
    \phi_{\rm s}&=(1-X_{\rm c})(1-\phi_{\rm p}),
    \end{split}
  \end{equation}
  The multiple chain free energy expression  can then be rewritten in the following way,
  \begin{widetext}
\begin{equation}\label{eq:appfree}
  \begin{split}
    f&=\frac{\phi_{\rm p}}{N}\ln{\phi_{\rm p}}+X_{\rm c}(1-\phi_{\rm p})\ln{\left[X_{\rm c}(1-\phi_{\rm p})\right]}+(1-X_{\rm c})(1-\phi_{\rm p})\ln{\left[(1-X_{\rm c})(1-\phi_{\rm p})\right]}\\ &+\chi_{\rm cs}(1-X_{\rm c})X_{\rm c}(1-\phi_{\rm p})^{2}+\chi_{\rm ps}(1-X_{\rm c})\phi_{\rm p}(1-\phi_{\rm p})+\chi_{\rm pc}X_{\rm c}(1-\phi_{\rm p})\phi_{\rm p}.
 \end{split}
\end{equation}
\end{widetext}
Here, the second virial coefficient $B$ is the coefficient of $\phi_{\rm p}^{2}$ and has the following form,
\begin{equation}\label{eq:appovirial}
\begin{split}
 B&=1-[2(1-X_{\rm c})\chi_{\rm ps} + 2X_{\rm c}\chi_{\rm pc} - 2X_{\rm c}(1-X_{\rm c})\chi_{\rm cs}].
\end{split}
\end{equation}
In the case of low concentration, the virial coefficient is
\begin{equation}
\begin{split}
B&=1-\left(2\chi_{\rm ps} + 2X_{\rm c}\chi_{\rm pc} - 2X_{\rm c}\chi_{\rm cs}\right),\\
&=1-2\left[\chi_{\rm ps} +X_{\rm c}(\chi_{\rm pc} - \chi_{\rm cs})\right].
\end{split}
\end{equation}

\newpage
\section{Coil-to-globule transition}
\label{sec:ctog}
The swelling parameter can be expressed in terms of the overall volume fractions in the following way,
\begin{equation}
\begin{split}
\phi_{\rm p}^{\rm '}&=\frac{N\sigma^{2}l}{V_{\rm \textsc{p}}}=\frac{3N\sigma^{2}}{4\pi R_{\rm g}^{3}}=\frac{3N\sigma^{2}}{4\pi\alpha^{3}N^{3/2}l^{2}}=\frac{k}{\sqrt{N}\alpha^{3}},\\
\alpha&=\left(\frac{k}{\phi_{\rm p}^{\rm '}\sqrt{N}}\right)^{1/3}=\left(\frac{\Phi^{\rm '}k}{\Phi_{\rm p}^{\rm '}\sqrt{N}}\right)^{1/3}=\left(\frac{\Phi^{\rm '}k}{\Phi_{\rm p}^{\rm '}\sqrt{N}}\right)^{1/3},
\end{split}
\end{equation}
where
\begin{equation}\label{eq:rigidity}
\kappa\sim \left(\frac{\sigma}{l}\right)^{2}.
\end{equation}
The overall volume of the system is  constant and can be expressed in terms of the molecular weight ($N$) and the overall polymer volume fraction ($\Phi_{\rm p}$) in the following way,
\begin{equation}
\frac{V}{v}=\frac{N}{\Phi_{\rm p}}.
\end{equation}
\newpage
\section{Mean energy of the bulk solvent-cosolvent mixture}
\label{sec:epsilonbulk}
Let the number of solvent and cosolvent in the bulk be $N^{\rm bulk}_{\rm s}$ and $N^{\rm bulk}_{\rm c}$, respectively. The fraction of the cosolvent in the bulk by the following expression,
\begin{equation}
X^{\rm bulk}_{\rm c}=\frac{N^{\rm bulk}_{\rm c}}{N^{\rm bulk}_{\rm c}+N^{\rm bulk}_{\rm s}}.
\end{equation}
The total mean energy of the bulk solvent-cosolvent mixture can be evaluated in the following way,
\begin{equation}
\begin{split}
\overline{E}_{\rm bulk}&=N^{\rm bulk}_{\rm c}\left(X^{\rm bulk}_{\rm c}\overline{\epsilon}_{\rm \textsc{cc}}+(1-X^{\rm bulk}_{\rm c})\overline{\epsilon}_{\rm \textsc{cs}}\right)+N^{\rm bulk}_{\rm s}\left((1-X^{\rm bulk}_{\rm c})\overline{\epsilon}_{\rm \textsc{ss}}+X^{\rm bulk}_{\rm c}\overline{\epsilon}_{\rm \textsc{cs}}\right)\\
&=N^{\rm bulk}_{\rm c}X^{\rm bulk}_{\rm c}\overline{\epsilon}_{\rm \textsc{cc}}+N^{\rm bulk}_{\rm s}(1-X^{\rm bulk}_{\rm c})\overline{\epsilon}_{\rm \textsc{ss}}+\overline{\epsilon}_{\rm \textsc{cs}}(N^{\rm bulk}_{\rm s}X^{\rm bulk}_{\rm c}+N^{\rm bulk}_{\rm c}(1-X^{\rm bulk}_{\rm c}))\\
&=(N^{\rm bulk}_{\rm c}+N^{\rm bulk}_{\rm s})((X^{\rm bulk}_{\rm c})^{2}\ \overline{\epsilon}_{\rm \textsc{cc}}+(1-X^{\rm bulk}_{\rm c})^{2}\ \overline{\epsilon}_{\rm \textsc{cc}}+2X^{\rm bulk}_{\rm c}(1-X^{\rm bulk}_{\rm c})\overline{\epsilon}_{\rm \textsc{cc}}),
\end{split}
\end{equation}
and the average mean energy of the mixture is the following,
\begin{equation}
\overline{\epsilon}_{\rm bulk}=\frac{\overline{E}_{\rm bulk}}{N^{\rm bulk}_{\rm c}+N^{\rm bulk}_{\rm s}}=(X^{\rm bulk}_{\rm c})^{2}\ \overline{\epsilon}_{\rm \textsc{cc}}+(1-X^{\rm bulk}_{\rm c})^{2}\ \overline{\epsilon}_{\rm \textsc{cc}}+2X^{\rm bulk}_{\rm c}(1-X^{\rm bulk}_{\rm c})\overline{\epsilon}_{\rm \textsc{cc}}.
\end{equation}
Given the dilute nature of the system, we have $X_{\rm c}^{\rm bulk}=X_{\rm c}$.
\newpage
\section{Simulation master curves}
The figure shows the master curves for $\epsilon_{\rm \textsc{as}}=1.7,1.8$. 
\begin{figure}[h] 
\begin{center}    
\subfigure{\label{fig:rt1}\includegraphics[scale=0.5]{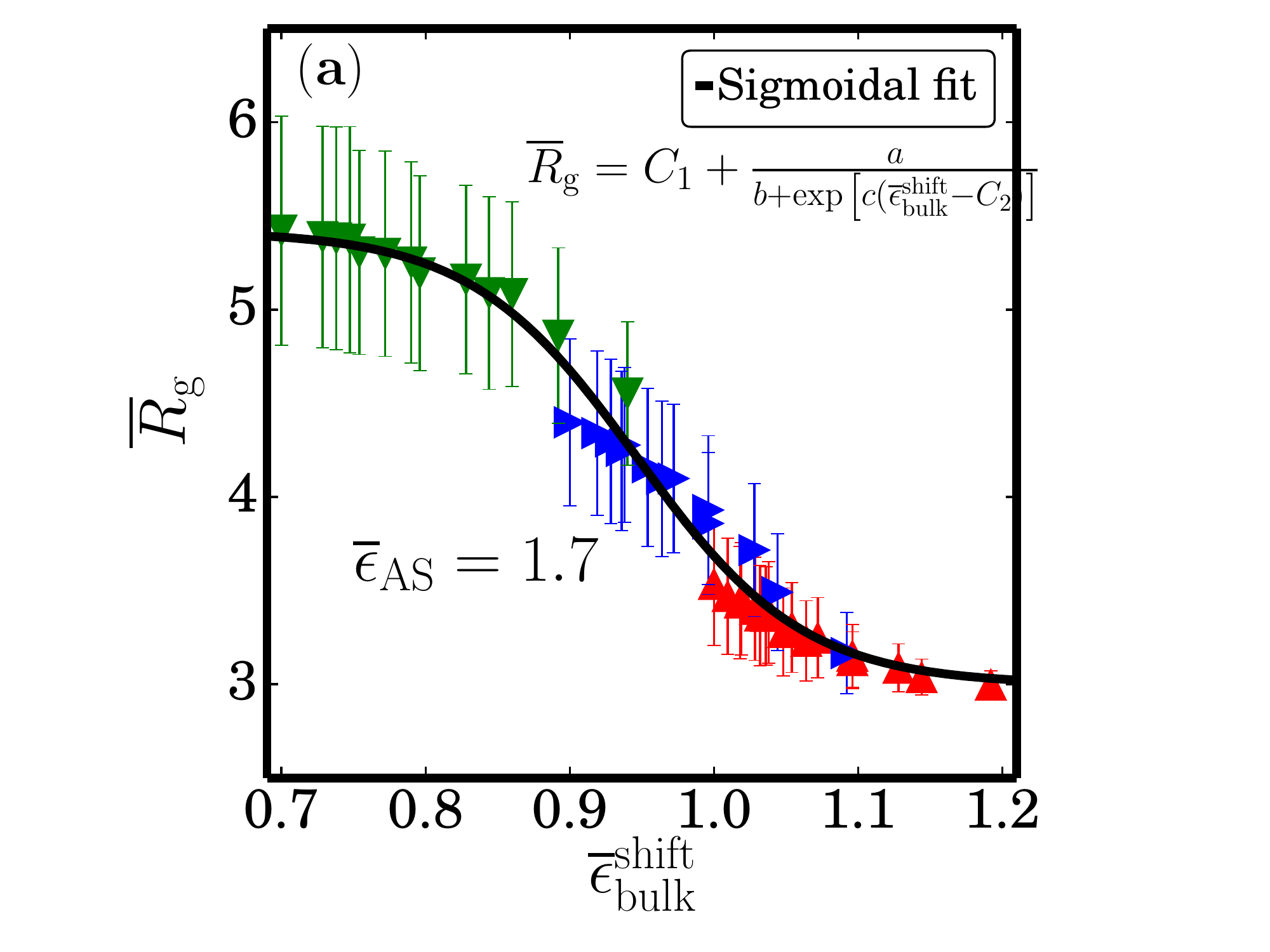}}
\subfigure{\label{fig:rt2}\includegraphics[scale=0.5]{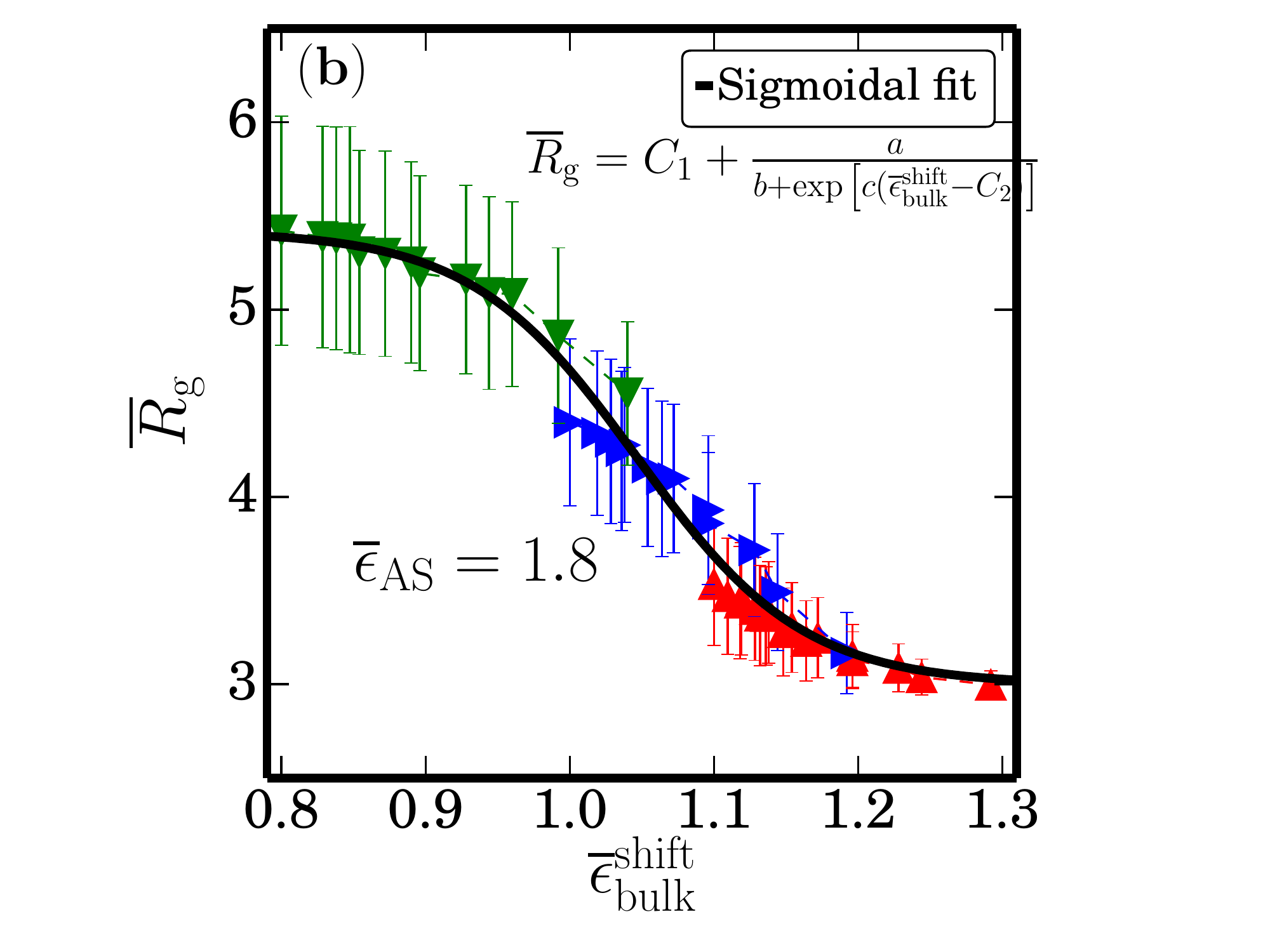}}
\end{center}
\caption{Variation of $\overline{R}_{\rm g}$ with $\overline{\epsilon}_{\rm bulk}^{\rm shift}=\overline{\epsilon}_{\rm bulk}+\Delta \overline{\epsilon}_{\rm bulk}$ master curves and corresponding sigmoidal fits of the form $\overline{R}_{\rm g}=C_{\rm 1}+a/[b+\exp{\left(c(\overline{\epsilon}_{\rm bulk}^{\rm shift}-C_{2})\right)}]$  (a)for $\overline{\epsilon}_{\rm \textsc{as}}=1.7$ where $a=0.96, b=0.39, c=17.4, C_{1}=3.0, C_{2}=1.0$ (b) for $\overline{\epsilon}_{\rm \textsc{as}}=1.8$ where $a=0.96, b=0.39, c=17.4, C_{1}=3.0, C_{2}=1.1$. }
\end{figure}

\end{document}